%% file: main.tex
\documentclass[12pt,a4paper]{article}

\usepackage{ifthen} 
\newboolean{pdflatex}
\setboolean{pdflatex}{false} 

\newboolean{articletitles}
\setboolean{articletitles}{true} 

\newboolean{uprightparticles}
\setboolean{uprightparticles}{false} 

\newboolean{inbibliography}
\setboolean{inbibliography}{false} 

\input{preamble}

\usepackage{booktabs} 
\definecolor{orange}{rgb}{1,0.5,0}

\usepackage{multirow}

\begin{document}

\renewcommand{\thefootnote}{\fnsymbol{footnote}}
\setcounter{footnote}{1}

\input{title-LHCb-PAPER}


\renewcommand{\thefootnote}{\arabic{footnote}}
\setcounter{footnote}{0}



\pagestyle{plain} 
\setcounter{page}{1}
\pagenumbering{arabic}


\input{introduction}

\input{LHCbdetector}

\input{selection}

\input{efficiency}

\input{fit}

\input{systematics}

\input{interpretation}

\input{conclusion}

\input{acknowledgements}

\addcontentsline{toc}{section}{References}
\setboolean{inbibliography}{true}
\bibliographystyle{LHCb}
\bibliography{main,LHCb-PAPER,LHCb-CONF,LHCb-DP,LHCb-TDR,mybiblio}

\newpage


\input{LHCb_Authorship_flat_16-Feb-2016}

\end{document}

%% file: preamble.tex
\usepackage[utf8]{inputenc}
\usepackage[T1]{fontenc}

\usepackage{microtype}
\usepackage{lineno}  
\usepackage{xspace} 
\usepackage{caption} 

\usepackage{graphicx}  
\usepackage{color}
\usepackage{colortbl}
\graphicspath{{./figs/}} 

\usepackage{amsmath} 
\usepackage{amssymb}
\usepackage{amsfonts}
\usepackage{upgreek} 

\newcommand*\patchAmsMathEnvironmentForLineno[1]{%
\expandafter\let\csname old#1\expandafter\endcsname\csname #1\endcsname
\expandafter\let\csname oldend#1\expandafter\endcsname\csname
end#1\endcsname
 \renewenvironment{#1}%
   {\linenomath\csname old#1\endcsname}%
   {\csname oldend#1\endcsname\endlinenomath}%
}
\newcommand*\patchBothAmsMathEnvironmentsForLineno[1]{%
  \patchAmsMathEnvironmentForLineno{#1}%
  \patchAmsMathEnvironmentForLineno{#1*}%
}
\AtBeginDocument{%
\patchBothAmsMathEnvironmentsForLineno{equation}%
\patchBothAmsMathEnvironmentsForLineno{align}%
\patchBothAmsMathEnvironmentsForLineno{flalign}%
\patchBothAmsMathEnvironmentsForLineno{alignat}%
\patchBothAmsMathEnvironmentsForLineno{gather}%
\patchBothAmsMathEnvironmentsForLineno{multline}%
\patchBothAmsMathEnvironmentsForLineno{eqnarray}%
}

\usepackage{hyperref}    
\usepackage[all]{hypcap} 

\input{lhcb-symbols-def} 
\input{my-symbols-def}

\usepackage[lofdepth,lotdepth]{subfig}

\usepackage{cite} 
\usepackage{mciteplus}

%% file: lhcb-symbols-def.tex
\usepackage{xspace} 
\usepackage{upgreek}

\def\lhcb {\mbox{LHCb}\xspace}

\def\babar  {\mbox{BaBar}\xspace}
\def\belle  {\mbox{Belle}\xspace}

\def\MagUp {\mbox{\em Mag\kern -0.05em Up}\xspace}

\ifthenelse{\boolean{uprightparticles}}%
{

 \def\Ppi         {\ensuremath{\uppi}\xspace}                 
                  
 \def\Prho        {\ensuremath{\uprho}\xspace}

 \def\PDelta      {\ensuremath{\Delta}\xspace}                 
 \def\PXi      {\ensuremath{\Xi}\xspace}                 
 \def\PLambda      {\ensuremath{\Lambda}\xspace}                 
 \def\PSigma      {\ensuremath{\Sigma}\xspace}                 
 \def\POmega      {\ensuremath{\Omega}\xspace}                 
 \def\PUpsilon      {\ensuremath{\Upsilon}\xspace}                 
 
 \def\PB      {\ensuremath{\mathrm{B}}\xspace}                 
                  
 \def\PD      {\ensuremath{\mathrm{D}}\xspace}

 \def\PK      {\ensuremath{\mathrm{K}}\xspace}

 \def\PX      {\ensuremath{\mathrm{X}}\xspace}

 \def\Pb      {\ensuremath{\mathrm{b}}\xspace}                 
 \def\Pc      {\ensuremath{\mathrm{c}}\xspace}                 
 \def\Pd      {\ensuremath{\mathrm{d}}\xspace}

 \def\Pi      {\ensuremath{\mathrm{i}}\xspace}

 \def\Pp      {\ensuremath{\mathrm{p}}\xspace}

 \def\Ps      {\ensuremath{\mathrm{s}}\xspace}                 
                  
 \def\Pu      {\ensuremath{\mathrm{u}}\xspace}

}
{

 \def\Ppi         {\ensuremath{\pi}\xspace}                 
                  
 \def\Prho        {\ensuremath{\rho}\xspace}

 \mathchardef\PDelta="7101
 \mathchardef\PXi="7104
 \mathchardef\PLambda="7103
 \mathchardef\PSigma="7106
 \mathchardef\POmega="710A
 \mathchardef\PUpsilon="7107
                  
 \def\PB      {\ensuremath{B}\xspace}                 
                  
 \def\PD      {\ensuremath{D}\xspace}

 \def\PK      {\ensuremath{K}\xspace}

 \def\PX      {\ensuremath{X}\xspace}

 \def\Pb      {\ensuremath{b}\xspace}                 
 \def\Pc      {\ensuremath{c}\xspace}                 
 \def\Pd      {\ensuremath{d}\xspace}

 \def\Pi      {\ensuremath{i}\xspace}

 \def\Pp      {\ensuremath{p}\xspace}

 \def\Ps      {\ensuremath{s}\xspace}                 
                  
 \def\Pu      {\ensuremath{u}\xspace}

}

\makeatletter
\ifcase \@ptsize \relax
  \newcommand{\miniscule}{\@setfontsize\miniscule{4}{5}}
\or
  \newcommand{\miniscule}{\@setfontsize\miniscule{5}{6}}
\or
  \newcommand{\miniscule}{\@setfontsize\miniscule{5}{6}}
\fi
\makeatother

\DeclareRobustCommand{\optbar}[1]{\shortstack{{\miniscule (\rule[.5ex]{1.25em}{.18mm})}
  \\ [-.7ex] $#1$}}

\def\uquark    {{\ensuremath{\Pu}}\xspace}

\def\dquark    {{\ensuremath{\Pd}}\xspace}

\def\squark    {{\ensuremath{\Ps}}\xspace}

\def\cquark    {{\ensuremath{\Pc}}\xspace}

\def\bquark    {{\ensuremath{\Pb}}\xspace}

\def\pion   {{\ensuremath{\Ppi}}\xspace}
\def\piz    {{\ensuremath{\pion^0}}\xspace}

\def\pip    {{\ensuremath{\pion^+}}\xspace}
\def\pim    {{\ensuremath{\pion^-}}\xspace}
\def\pipm   {{\ensuremath{\pion^\pm}}\xspace}
\def\pimp   {{\ensuremath{\pion^\mp}}\xspace}

\def\rhomeson {{\ensuremath{\Prho}}\xspace}
\def\rhoz     {{\ensuremath{\rhomeson^0}}\xspace}

\def\kaon    {{\ensuremath{\PK}}\xspace}
  \def\Kbar    {{\kern 0.2em\overline{\kern -0.2em \PK}{}}\xspace}

\def\KorKbar    {\kern 0.18em\optbar{\kern -0.18em K}{}\xspace}

\def\Kp      {{\ensuremath{\kaon^+}}\xspace}
\def\Km      {{\ensuremath{\kaon^-}}\xspace}
\def\Kpm     {{\ensuremath{\kaon^\pm}}\xspace}
\def\Kmp     {{\ensuremath{\kaon^\mp}}\xspace}
\def\KS      {{\ensuremath{\kaon^0_{\mathrm{ \scriptscriptstyle S}}}}\xspace}

\def\Kstarz  {{\ensuremath{\kaon^{*0}}}\xspace}
\def\Kstarzb {{\ensuremath{\Kbar{}^{*0}}}\xspace}
\def\Kstar   {{\ensuremath{\kaon^*}}\xspace}

  \def\Dbar    {{\kern 0.2em\overline{\kern -0.2em \PD}{}}\xspace}
\def\D       {{\ensuremath{\PD}}\xspace}

\def\DorDbar    {\kern 0.18em\optbar{\kern -0.18em D}{}\xspace}
\def\Dz      {{\ensuremath{\D^0}}\xspace}
\def\Dzb     {{\ensuremath{\Dbar{}^0}}\xspace}

\def\Dpm     {{\ensuremath{\D^\pm}}\xspace}

\def\Dstar   {{\ensuremath{\D^*}}\xspace}

\def\Dstarz  {{\ensuremath{\D^{*0}}}\xspace}
\def\Dstarzb {{\ensuremath{\Dbar{}^{*0}}}\xspace}
\def\Dstarp  {{\ensuremath{\D^{*+}}}\xspace}

\def\B       {{\ensuremath{\PB}}\xspace}
\def\Bbar    {{\ensuremath{\kern 0.18em\overline{\kern -0.18em \PB}{}}}\xspace}

\def\BorBbar    {\kern 0.18em\optbar{\kern -0.18em B}{}\xspace}
\def\Bz      {{\ensuremath{\B^0}}\xspace}
\def\Bzb     {{\ensuremath{\Bbar{}^0}}\xspace}
\def\Bu      {{\ensuremath{\B^+}}\xspace}

\def\Bp      {{\ensuremath{\Bu}}\xspace}

\def\Bpm     {{\ensuremath{\B^\pm}}\xspace}

\def\Bd      {{\ensuremath{\B^0}}\xspace}
\def\Bs      {{\ensuremath{\B^0_\squark}}\xspace}
\def\Bsb     {{\ensuremath{\Bbar{}^0_\squark}}\xspace}
\def\Bdb     {{\ensuremath{\Bbar{}^0}}\xspace}

  \def\Y#1S{\ensuremath{\PUpsilon{(#1S)}}\xspace}

\def\proton      {{\ensuremath{\Pp}}\xspace}

\def\Lz          {{\ensuremath{\PLambda}}\xspace}
\def\Lbar        {{\ensuremath{\kern 0.1em\overline{\kern -0.1em\PLambda}}}\xspace}
\def\LorLbar    {\kern 0.18em\optbar{\kern -0.18em \PLambda}{}\xspace}

\def\Lb      {{\ensuremath{\Lz^0_\bquark}}\xspace}

\def\BF         {{\ensuremath{\mathcal{B}}}\xspace}

\def\BR         {\BF}
\newcommand{\decay}[2]{\ensuremath{#1\!\to #2}\xspace}         

\def\to                 {\ensuremath{\rightarrow}\xspace}

\newcommand{\mBd}{{\ensuremath{m_{\Bd}}}\xspace}

\def\order   {{\ensuremath{\mathcal{O}}}\xspace}

\def\eps   {{\ensuremath{\varepsilon}}\xspace}

\def\CP                {{\ensuremath{C\!P}}\xspace}

\def\Vud  {{\ensuremath{V_{\uquark\dquark}}}\xspace}
\def\Vcd  {{\ensuremath{V_{\cquark\dquark}}}\xspace}

\def\Vub  {{\ensuremath{V_{\uquark\bquark}}}\xspace}
\def\Vcb  {{\ensuremath{V_{\cquark\bquark}}}\xspace}

\def\AT#1     {\ensuremath{A_{\mathrm{T}}^{#1}}\xspace}           

\def\C#1      {\ensuremath{\mathcal{C}_{#1}}\xspace}                       
\def\Cp#1     {\ensuremath{\mathcal{C}_{#1}^{'}}\xspace}                    
\def\Ceff#1   {\ensuremath{\mathcal{C}_{#1}^{\mathrm{(eff)}}}\xspace}        
\def\Cpeff#1  {\ensuremath{\mathcal{C}_{#1}^{'\mathrm{(eff)}}}\xspace}       
\def\Ope#1    {\ensuremath{\mathcal{O}_{#1}}\xspace}                       
\def\Opep#1   {\ensuremath{\mathcal{O}_{#1}^{'}}\xspace}                    

\newcommand{\tev}{\ifthenelse{\boolean{inbibliography}}{\ensuremath{~T\kern -0.05em eV}\xspace}{\ensuremath{\mathrm{\,Te\kern -0.1em V}}}\xspace}
\newcommand{\gev}{\ensuremath{\mathrm{\,Ge\kern -0.1em V}}\xspace}
\newcommand{\mev}{\ensuremath{\mathrm{\,Me\kern -0.1em V}}\xspace}
\newcommand{\kev}{\ensuremath{\mathrm{\,ke\kern -0.1em V}}\xspace}
\newcommand{\ev}{\ensuremath{\mathrm{\,e\kern -0.1em V}}\xspace}
\newcommand{\gevc}{\ensuremath{{\mathrm{\,Ge\kern -0.1em V\!}}}\xspace}
\newcommand{\mevc}{\ensuremath{{\mathrm{\,Me\kern -0.1em V\!}}}\xspace}
\newcommand{\gevcc}{\ensuremath{{\mathrm{\,Ge\kern -0.1em V\!}}}\xspace}
\newcommand{\gevgevcccc}{\ensuremath{{\mathrm{\,Ge\kern -0.1em V^2\!}}}\xspace}
\newcommand{\mevcc}{\ensuremath{{\mathrm{\,Me\kern -0.1em V\!}}}\xspace}

\def\mum  {\ensuremath{{\,\upmu\mathrm{m}}}\xspace}

\def\invfb   {\ensuremath{\mbox{\,fb}^{-1}}\xspace}

\def\order{{\ensuremath{\mathcal{O}}}\xspace}
\newcommand{\chisq}{\ensuremath{\chi^2}\xspace}

\newcommand{\chisqip}{\ensuremath{\chi^2_{\text{IP}}}\xspace}

\def\deriv {\ensuremath{\mathrm{d}}}

\def\gsim{{~\raise.15em\hbox{$>$}\kern-.85em
          \lower.35em\hbox{$\sim$}~}\xspace}
\def\lsim{{~\raise.15em\hbox{$<$}\kern-.85em
          \lower.35em\hbox{$\sim$}~}\xspace}

\newcommand{\Real}{\ensuremath{\mathcal{R}e}\xspace}
\newcommand{\Imag}{\ensuremath{\mathcal{I}m}\xspace}

\def\sqs   {\ensuremath{\protect\sqrt{s}}\xspace}

\def\ptot       {\mbox{$p$}\xspace}
\def\pt         {\mbox{$p_{\mathrm{ T}}$}\xspace}

\def\degrees{\ensuremath{^{\circ}}\xspace}

\def\evtgen     {\mbox{\textsc{EvtGen}}\xspace}

\def\geant      {\mbox{\textsc{Geant4}}\xspace}

\def\photos     {\mbox{\textsc{Photos}}\xspace}

\def\pythia     {\mbox{\textsc{Pythia}}\xspace}

\def\tell1  {TELL1\xspace}
\def\ukl1   {UKL1\xspace}

\newcommand{\eg}{\mbox{\itshape e.g.}\xspace}

%% file: my-symbols-def.tex
\def\cosT    {{\ensuremath{\cos{\theta^*}}}\xspace}

\def\BdbDKstb      {\decay{\Bdb}{\D\Kstarzb}}
\def\BdDKst        {\decay{\Bd}{\D\Kstarz}}

\def\BdDKstKSpipi  {\decay{\Bd}{\D(\KS\pip\pim)\Kstarz}}
\def\DKSpipi       {\decay{\D}{\KS\pip\pim}}
\def\BDK        {\decay{\Bpm}{\D\Kpm}}

\def\BdDKPi        {\decay{\Bd}{\D\Kp\pim}}

\def\BsDKst        {\decay{\Bs}{\D\Kstarzb}}

\def\BdDrho        {\decay{\Bd}{\D\rhoz}}

\def\BuDpipipi        {\decay{\Bu}{\Dzb\pip\pip\pim}}

\def\BsDstKst     {\decay{\Bs}{\Dstar\Kstarzb}}

\def\BsDstKstpiz  {\decay{\Bs}{\Dstar(\Dz\piz)\Kstarzb}}
\def\BsDstKstgam  {\decay{\Bs}{\Dstar(\Dz\gam)\Kstarzb}}

\def\BdDstKst     {\decay{\Bd}{\Dstar\Kstarz}}

\def\KstzorKstzb {{\ensuremath{\KorKbar{}^{*0}}}\xspace}

\def\BdsDstKst     {\decay{\Bdors}{\Dstar\KstzorKstzb}}

\def\Bdors   {{\ensuremath{\B^0_{(\squark)}}}\xspace}
\def\BdsDKst {\decay{\Bdors}{\D\KstzorKstzb}}
\def\BdsDKst {\decay{\Bdors}{\D\KstzorKstzb}}

\def\DstDpiz {\decay{\Dstarz}{\Dz\piz}}
\def\DstDgam {\decay{\Dstarz}{\Dz\gam}}

\def\Dfourpi {\decay{\Dz}{\pi\pi\pi\pi}}

\def\LbDppi  {\decay{\Lb}{\Dz\proton\pim}}

\def\gam   {{\ensuremath{\gamma}}\xspace}
\def\xp   {{\ensuremath{x_{+}}}\xspace}
\def\xm   {{\ensuremath{x_{-}}}\xspace}
\def\yp   {{\ensuremath{y_{+}}}\xspace}
\def\ym   {{\ensuremath{y_{-}}}\xspace}
\def\xy   {{\ensuremath{(x_{\pm},y_{\pm})}}\xspace}
\def\zpm   {{\ensuremath{z_{\pm}}}\xspace}
\def\xpm   {{\ensuremath{x_{\pm}}}\xspace}
\def\ypm   {{\ensuremath{y_{\pm}}}\xspace}

\def\rbz {{\ensuremath{r_{\Bz}}}\xspace}
\def\rbzSq {{\ensuremath{r_{\Bz}^2}}\xspace}

\def\deltabz {{\ensuremath{\delta_{\Bz}}}\xspace}
\def\kap {{\ensuremath{\kappa}}\xspace}
\def\rbzdbzgam {{\ensuremath{(\rbz,\deltabz, \gamma)}}\xspace}

\def\dxm   {{\ensuremath{\delta x_{-}}}\xspace}
\def\dxp   {{\ensuremath{\delta x_{+}}}\xspace}
\def\dym   {{\ensuremath{\delta y_{-}}}\xspace}
\def\dyp   {{\ensuremath{\delta y_{+}}}\xspace}

\def\X {{\ensuremath{\PX}}\xspace}
\def\Xsz {{\ensuremath{\X{}^0_\squark}}\xspace}
\def\Xbar    {{\kern 0.2em\overline{\kern -0.2em \PX}{}}\xspace}
\def\Xszb     {{\ensuremath{\Xbar{}^0_\squark}}\xspace}

\def\Kst {{\ensuremath{\Kstar(892)^0}}\xspace}
\def\Kstpm {{\ensuremath{\Kstar(892)^\pm}}\xspace}

\def\KSpipi {\KS\pip\pim}

\def\mSqp {{\ensuremath{m^2_+}}\xspace}
\def\mSqm {{\ensuremath{m^2_-}}\xspace}
\def\mSqz {{\ensuremath{m^2_0}}\xspace}

\def\mSqpm {{\ensuremath{m^2_\pm}}\xspace}

%% file: title-LHCb-PAPER.tex

\begin{titlepage}
\pagenumbering{roman}

\vspace*{-1.5cm}
\centerline{\large EUROPEAN ORGANIZATION FOR NUCLEAR RESEARCH (CERN)}
\vspace*{1.5cm}
\noindent
\begin{tabular*}{\linewidth}{lc@{\extracolsep{\fill}}r@{\extracolsep{0pt}}}
\ifthenelse{\boolean{pdflatex}}
{\vspace*{-2.7cm}\mbox{\!\!\!\includegraphics[width=.14\textwidth]{lhcb-logo.pdf}} & &}%
{\vspace*{-1.2cm}\mbox{\!\!\!\includegraphics[width=.12\textwidth]{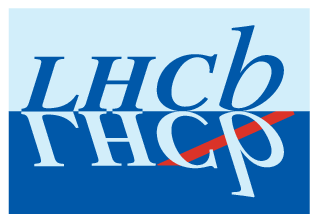}} & &}%
\\
 & & CERN-EP-2016-089 \\  
 & & LHCb-PAPER-2016-007 \\  
 & & May 3, 2016  \\ 
 & & \\
\end{tabular*}

\vspace*{.5cm}

{\normalfont\bfseries\boldmath\huge
\begin{center}
Measurement of the CKM angle \gam using \BdDKst with \DKSpipi decays
\end{center}
}

\vspace*{.5cm}

\begin{center}
The LHCb collaboration\footnote{Authors are listed at the end of this paper.}
\end{center}

\vspace{\fill}

\begin{abstract}
  \noindent
A model-dependent amplitude analysis of the decay \BdDKstKSpipi is performed using proton-proton collision data 
corresponding to an integrated luminosity of 3.0\invfb,
recorded at $\sqs=7$ and $8\tev$ by the LHCb experiment. The \CP violation observables $x_{\pm}$ and $y_{\pm}$,
sensitive to the CKM angle \gam, are measured to be
\begin{eqnarray*}
x_- &=& -0.15 \pm 0.14  \pm 0.03 \pm 0.01,\\
y_- &=& \phantom{-}0.25 \pm 0.15 \pm 0.06 \pm 0.01,\\
x_+ &=& \phantom{-}0.05 \pm 0.24 \pm 0.04 \pm 0.01,\\
y_+ &=& -0.65~^{+0.24~~}_{-0.23~~} \pm 0.08 \pm 0.01,
\end{eqnarray*}
where the first uncertainties are statistical, the second systematic and the third arise from the uncertainty on the
\DKSpipi amplitude model. These are the most precise measurements of these observables. They correspond to $\gamma=(80^{+21}_{-22})^{\circ}$ and $\rbz=0.39\pm0.13$, where \rbz is the magnitude of the ratio of the suppressed and favoured \BdDKPi decay amplitudes, 
in a $K\pi$ mass region of $\pm50\mevcc$ around the \Kst mass and for an absolute value of the cosine of the 
\Kstarz decay angle larger than $0.4$.

\end{abstract}

\vspace*{.5cm}

\begin{center}
  Published in JHEP 08 (2016) 137
\end{center}

\vspace{\fill}

{\footnotesize 
\centerline{\copyright~CERN on behalf of the \lhcb collaboration, licence \href{http://creativecommons.org/licenses/by/4.0/}{CC-BY-4.0}.}}
\vspace*{2mm}

\end{titlepage}


\newpage
\setcounter{page}{2}
\mbox{~}
%
%
%
%

\cleardoublepage

%% file: introduction.tex
\section{Introduction}
\label{sec:Introduction}
\newcommand{\bl}{\hphantom{-}}
\newcommand{\bn}{\hphantom{9}}

The Standard Model can be tested by checking the consistency of the Cabibbo-Kobayashi-Maskawa (CKM) 
mechanism~\cite{Cabibbo,KM}, which describes the mixing between weak and mass
eigenstates of the quarks.
The CKM phase \gam can be expressed in terms of the elements of
the complex unitary CKM matrix, as $\gamma \equiv \mathrm{arg} \left[- \Vud \Vub^* / \Vcd \Vcb^*  \right]$.
Since \gam is also the angle of the unitarity triangle least constrained by direct measurements,
its precise determination is of considerable interest. Its value can be measured in 
tree-level processes such as \BDK and \BdDKst, where \D is a superposition of the \Dz\ and \Dzb flavour eigenstates,
 and \Kstarz is the \Kst meson. 
Since loop corrections to these processes are of higher order, the associated theoretical
uncertainty on $\gamma$ is negligible~\cite{GammaTheoError}.
As such, measurements of \gam in tree-level decays provide a reference value, 
allowing searches for potential deviations due to physics beyond the Standard Model in other processes.

The combination of measurements by the \babar~\cite{BaBar-Gamma-2013} and \belle~\cite{Belle-Gamma-2013}
 collaborations gives $\gam=\left(67\pm11\right)\degrees$~\cite{Bevan:2014iga}, 
whilst
an average value of \lhcb determinations in 2014 gave $\gam=\left(73^{+9}_{-10}\right)\degrees$~\cite{LHCb-CONF-2014-004}.
Global fits of all current CKM measurements by the CKMfitter~\cite{CKMFitter,CKMFitter2} and UTfit~\cite{UTFit}
collaborations yield indirect estimates of \gam with an uncertainty of $2^{\circ}$. Some of the CKM measurements
included in these combinations can be affected by new physics contributions.

Since the phase difference between \Vub and \Vcb depends on \gam, the determination of \gam in tree-level 
decays relies on the interference between $b \rightarrow c$ and $b \rightarrow u$ transitions.
The strategy of using \BDK decays to determine \gam from an amplitude analysis of \D-meson decays to the
three-body
final state \KSpipi
was first proposed in Refs.~\cite{GGSZ-Bondar,GGSZ}.  
The method requires knowledge of the \DKSpipi decay amplitude
across the phase space, and in particular the variation of its strong phase.
This may be obtained either by using a model to describe the $D$-meson decay 
amplitude in phase space (model-dependent approach), or by using measurements of the 
phase behaviour of the amplitude (model-independent approach). 
The model-independent strategy, used by \belle~\cite{Aihara:2012aw} 
and \lhcb~\cite{LHCb-PAPER-2014-041, LHCb-PAPER-2012-027},
incorporates measurements from CLEO~\cite{Libby:2010nu} of the \D\ decay strong phase in bins across the
phase space.
The present paper reports a new unbinned model-dependent measurement, following the method used by the 
\babar~\cite{Aubert:2005iz, Aubert:2008bd, delAmoSanchez:2010rq}, 
\belle~\cite{Poluektov:2004mf, Poluektov:2006ia, Poluektov:2010wz} and
\lhcb~\cite{LHCb-PAPER-2014-017} collaborations
in their analyses of $\B^\pm \rightarrow D ^{(*)} K ^{(*)\pm}$ decays.
This method allows the statistical power of the data to be fully exploited.

The sensitivity to \gam depends both on the yield of the sample analysed and on the magnitude of the ratio $r_\B$ of the suppressed
and favoured decay amplitudes in the relevant region of phase space. Due to colour suppression, the branching fraction
$\BR(\decay{\Bz}{\Dzb\Kstarz})=(4.2\pm0.6)\times10^{-5}$ is an order of magnitude smaller than that of the corresponding 
charged \B-meson decay mode, $\BR(\decay{\Bp}{\Dzb\Kp})=(3.70\pm0.17)\times10^{-4}$~\cite{PDG2014}.
However, this is partially compensated by an enhancement in \rbz, which was measured to be $r_{B^0} = 0.240 ^{+0.055}_{-0.048}$
in \BdDKst decays in which the \D is reconstructed in two-body final states~\cite{LHCb-PAPER-2014-028};
the charged decays have an average value of $r_\B = 0.097\pm0.006$\cite{CKMFitter,CKMFitter2}.
Model-dependent and independent determinations of \gam using \BdDKstKSpipi decays have already been performed by the 
\babar~\cite{PhysRevD.79.072003} and \belle~\cite{Negishi:2015vqa} collaborations, respectively.  
The model-independent approach has also been employed recently by \lhcb~\cite{LHCb-PAPER-2016-006}.
For these decays a time-independent \CP analysis is performed, as the \Kstarz is reconstructed in the self-tagging
mode \Kp\pim, where the charge of the kaon provides the flavour of the decaying neutral \B meson.

The \Kstarz meson is one of several possible states of the (\Kp\pim) system.
Letting \Xsz represent any such state, the $B$-meson decay amplitude 
to $D K^+ \pi^-$ may be expressed as a
superposition of favoured $b \rightarrow c$ and suppressed $b \rightarrow u$ contributions:
\begin{align}
\begin{aligned}
\mathcal{A}(\Bzb \rightarrow D \Xszb) &\propto |A_c|     {A}_f + |A_u| e^{i(\delta_{B^0} - \gamma)} \bar{A}_f, \\
\mathcal{A}(\Bz  \rightarrow D \Xsz ) &\propto |A_c| \bar{A}_f + |A_u| e^{i(\delta_{B^0} + \gamma)}     {A}_f,
 \label{eq.intro.Ap}
\end{aligned}
\end{align}
where
$|A_{c,u}|$ are the magnitudes
of the favoured and suppressed $B$-meson decay amplitudes, $\delta_{\B^0}$ is the strong phase difference between them, and $\gamma$
is the \CP-violating weak phase. The quantities $A_{c,u}$ and $\delta_{\B^0}$ depend on the position in the $\Bz \rightarrow D K^+ \pi^-$ phase space.
The amplitudes of the \Dz\ and \Dzb\ mesons decaying into the common final
state $f$, \mbox{$A_f \equiv \left\langle f \vphantom{\bar{D}^0} \right| \mathcal{H} \left| \vphantom{\bar{D}^0} \Dz \right\rangle$} and
\mbox{$\bar{A}_f \equiv \left\langle f \vphantom{\bar{D}^0} \right| \mathcal{H} \left| \Dzb \right\rangle$},
are functions of the \KSpipi final state, which can be completely specified by two squared invariant masses of pairs of the three final-state
particles, chosen to be
\mbox{$m_{+}^2 \equiv m_{\KS \pip}^2$} and \mbox{$m_{-}^2 \equiv m_{\KS \pim}^2$}. The other squared invariant mass is 
\mbox{$m_{0}^2 \equiv m_{\pip \pim}^2$}.
Making the assumption of no \CP violation in the $D$-meson decay, the amplitudes $A_f$ and $\bar{A}_f$ are
related by $\bar{A}_f( m^2_+, m^2_- ) = A_f( m^2_-, m^2_+ )$.

The amplitudes in Eq.~\ref{eq.intro.Ap} give rise to distributions of 
the form
\begin{align}
  \begin{aligned}
    \deriv\Gamma_{\Bdb} \propto & \;  |A_c|^2 |A_{f}|^2 + |A_u|^2 |\bar{A}_{f}|^2 + 2|A_c| |A_u|\ \Real\left[A_{f}^\star \bar{A}_{f} \, e^{i(\deltabz - \gam)} \right], \\
    \deriv\Gamma_{\Bd}  \propto & \;  |A_c|^2 |\bar{A}_{f}|^2 + |A_u|^2 |A_{f}|^2 + 2|A_c| |A_u|\ \Real\left[A_{f} \bar{A}_{f}^\star \, e^{i(\deltabz + \gam)} \right],
  \end{aligned}
    \label{eq:PDFgeneral}
\end{align}
which are functions of the position in the $\Bz \rightarrow D K^+ \pi^-$ phase space.
Integrating only over the region $\phi_{\Kstarz}$ of the $\Bz \rightarrow D K^+ \pi^-$ phase space in which the \Kstarz resonance
is dominant,
\begin{equation}
\rbzSq \equiv \frac{\int_{\phi_{\Kstarz}}{\deriv\phi\ |A_u|^2}}
                   {\int_{\phi_{\Kstarz}}{\deriv\phi\ |A_c|^2}}. \label{eq:defrb}
\end{equation}
The functional
\begin{equation}
  \mathcal{P}( A, z, \kap ) =                    \left| \vphantom{\bar{A}}      A  \right|^2 +
                              \left| z \right|^2 \left|                    \bar{A} \right|^2 +
                              2 \kap \Real\left[ z A^\star \bar{A} \right],
\label{eq:functionalP}
\end{equation}
describes the distribution within the phase space of the $D$-meson decay,
\begin{align}
\begin{aligned}
  \mathcal{P}_{\Bdb}(\mSqm,\mSqp) \propto \mathcal{P}(     {A}_f, z_-, \kap ), \\
  \mathcal{P}_{\Bd} (\mSqm,\mSqp) \propto \mathcal{P}( \bar{A}_f, z_+, \kap ),
\end{aligned}
  \label{eq:pdf-exp}
\end{align}
where the coherence factor \kap is a real constant $(0 \leq \kap \leq 1)$~\cite{Gronau:2002mu}
measured in Ref.~\cite{LHCb-PAPER-2015-059}, 
parameterising the fraction of the region $\phi_{\Kstarz}$ that is occupied by the \Kstarz resonance,
and the complex parameters $z_\pm$ are
\begin{equation}
  z_\pm = \rbz \, e^{i ( \deltabz \pm \gam )}.
\label{eq:zpm}
\end{equation}

\noindent
A direct determination of \rbz, $\delta_{\B^0}$ and \gam\ can lead to bias, when \rbz gets close to zero~\cite{Aubert:2005iz}. 
The Cartesian \CP violation observables, \mbox{$x_{\pm} = \Real({z_\pm})$} and
\mbox{$y_{\pm} = \Imag({z_\pm})$}, are therefore used instead.

This paper reports model-dependent Cartesian measurements of $z_\pm$ made
using \BdDKstKSpipi decays selected from $pp$ collision data, corresponding to an integrated luminosity of
$3\invfb$, recorded by \lhcb at centre-of-mass energies of $7\tev$ in 2011 and $8\tev$ in 2012.
The measured values of $z_\pm$ place constraints on the CKM angle \gam.
Throughout the paper, inclusion of charge conjugate processes is implied,
unless specified otherwise.

Section~\ref{sec:Detector} describes the \lhcb detector used to record the data, 
and the methods used to produce a realistic simulation of the data.
Section~\ref{sec:selection} outlines the procedure used to select candidate \BdDKstKSpipi\ decays, and 
Sec.~\ref{sec:efficiency} describes the determination of the selection efficiency across the phase space of the $D$-meson decay. 
Section~\ref{sec:fit} details the fitting procedure used to determine the values of the Cartesian \CP violation observables and 
Sec.~\ref{sec:systematics} describes the systematic uncertainties on these results. 
Section~\ref{sec:interpretation} presents the interpretation of the measured Cartesian \CP violation observables in terms of central values and 
confidence intervals for \rbz, $\delta_{\B^0}$ and \gam, before
Sec.~\ref{sec:conclu} concludes with a summary of the results obtained.

%% file: LHCbdetector.tex
\section{The \lhcb detector}
\label{sec:Detector}

The \lhcb detector~\cite{Alves:2008zz,LHCb-DP-2014-002} is a single-arm forward
spectrometer covering the \mbox{pseudorapidity} range $2<\eta <5$,
designed for the study of particles containing \bquark or \cquark
quarks. The detector includes a high-precision tracking system
consisting of a silicon-strip vertex detector surrounding the $pp$
interaction region, a large-area silicon-strip detector located
upstream of a dipole magnet of reversible polarity with a bending power of about
$4{\mathrm{\,Tm}}$, and three stations of silicon-strip detectors and straw
drift tubes placed downstream of the magnet.
The tracking system provides a measurement of the momentum \ptot of charged particles with
a relative uncertainty that varies from 0.5\% at low momentum to 1.0\% at 200\gevc.
The minimum distance of a track to a primary vertex (PV), the impact parameter (IP), is measured with a resolution of $(15+29/\pt)\mum$,
where \pt is the component of the momentum transverse to the beam, in\,\gevc.
Different types of charged hadrons are distinguished using information
from two ring-imaging Cherenkov detectors.
Photons, electrons and hadrons are identified by a calorimeter system consisting of
scintillating-pad and preshower detectors, an electromagnetic
calorimeter and a hadronic calorimeter. Muons are identified by a
system composed of alternating layers of iron and multiwire
proportional chambers.

The trigger consists of a
hardware stage, based on information from the calorimeter and muon
systems, followed by a software stage, in which all charged particles
with $\pt>500\,(300)\mev$ are reconstructed for 2011\,(2012) data.
The software trigger requires a two-, three- or four-track
secondary vertex with a large sum of the transverse momentum, \pt, of
the tracks and a significant displacement from the primary $pp$
interaction vertices. At least one track should have $\pt >
1.7\gevc$ and \chisqip with respect to any
primary interaction greater than 16, where \chisqip is defined as the
difference in \chisq of a given PV reconstructed with and
without the considered track.
A multivariate algorithm~\cite{BBDT} is used for
the identification of secondary vertices consistent with the decay
of a \bquark hadron.
In the offline selection, trigger signals are associated with reconstructed particles.
Selection requirements can therefore be made on the trigger selection itself
and on whether the decision was due to the signal candidate, other particles produced in the $pp$ collision, or a combination of both.

Decays of \decay{\KS}{\pip\pim} are reconstructed in two different categories:
the first involving \KS mesons that decay early enough for the
daughter pions to be reconstructed in the vertex detector, and the
second containing \KS that decay later such that track segments of the
pions cannot be formed in the vertex detector. These categories are
referred to as \emph{long} and \emph{downstream}, respectively. The
long category has better mass, momentum and vertex resolution than the
downstream category.

Large samples of simulated \BdsDKst decays and various background decays are used in this study.
In the simulation, $pp$ collisions are generated using
\pythia~\cite{Sjostrand:2007gs,*Sjostrand:2006za}
 with a specific \lhcb
configuration~\cite{LHCb-PROC-2010-056}.  Decays of hadronic particles
are described by \evtgen~\cite{Lange:2001uf}, in which final-state
radiation is generated using \photos~\cite{Golonka:2005pn}. The
interaction of the generated particles with the detector, and its response,
are implemented using the \geant
toolkit~\cite{Allison:2006ve, *Agostinelli:2002hh}, as described in
Ref.~\cite{LHCb-PROC-2011-006}.

%% file: selection.tex
\section{Candidate selection and background sources}
\label{sec:selection}

In addition to the hardware and software trigger requirements, after a kinematic fit~\cite{Hulsbergen:2005pu} to constrain the \Bd candidate 
to point towards the PV and the \D candidate to have its nominal mass, the invariant mass of the \KS 
candidates must lie within $\pm 14.4\mev$ ($\pm 19.9\mev$) of the known value~\cite{PDG2014} for 
long (downstream) categories. Likewise, after a kinematic fit to constrain the \Bd candidate 
to point towards the PV and the \KS candidate to have the \KS mass,
the reconstructed \D-meson candidate must lie within $\pm 30\mev$ of the \Dz mass. 
To reconstruct the \Bd mass, a third kinematic fit of the whole decay chain is used, constraining the \Bd candidate 
 to point towards the PV and the \D and \KS to have their nominal masses. 
The \chisq of this fit is used in the multivariate classifier described below.
This fit improves the resolution of the \mSqpm invariant masses and
ensures that the reconstructed \D candidates are constrained to lie within the kinematic boundaries of the phase space.
The \Kstarz 
candidate must have a mass within $\pm50\mev$ of the world average value and $|\cosT|>0.4$, where the decay angle $\theta^*$ is defined in the \Kstarz rest frame
as the angle between the momentum of the kaon daughter of the \Kstarz, and the direction opposite to the \Bd momentum.
The criteria placed on the \Kstarz 
candidate are identical to those used in the analysis of \BdDKst with two-body \D decays~\cite{LHCb-PAPER-2014-028}. 

A multivariate classifier is then used to improve the signal purity. A boosted decision 
tree (BDT)~\cite{Breiman,Roe} is trained on simulated signal events and background candidates lying in
the high \Bd mass sideband $[5500,6000]\mev$ in data. This mass range partially overlaps with
the range of the invariant mass fit described below. To avoid a potential fit bias, the candidates
are randomly split into two disjoint subsamples, A and B, and two independent BDTs (BDTA and BDTB) are trained with them. 
These classifiers are then applied to the complementary samples. The BDTs are based on 16
discriminating variables: the \Bd meson \chisqip, the sum of the \chisqip of the \KS daughter pions,
the sum of the \chisqip of the final state particles except the \KS daughters, the \Bd and \D
decay vertex \chisq, the values of the flight distance significance with respect to the PV for the \Bd, \D and \KS mesons,
the \D 
(\KS) flight distance significance with respect to the \Bd (\D) decay vertex, the transverse momenta of the \Bd, \D and \Kstarz, 
the cosine of the angle between the momentum direction of the \Bd and the displacement vector from 
the PV to the \Bd decay vertex, the decay angle of the \Kstarz and the \chisq of the 
kinematic fit of the whole decay chain.
Since some of the variables have different distributions for 
long or downstream candidates, the two event categories have separate BDTs, giving a total 
of four independent BDTs.
The optimal cut value of each BDT classifier is chosen from pseudoexperiments to minimise the
uncertainties on \zpm.

Particle identification (PID) requirements are applied to the daughters of the \Kstarz to select 
kaon-pion pairs and reduce background coming from \BdDrho decays. A specific veto is also
applied to remove contributions from \BDK decays: \BdDKst candidates with a \D\kaon 
invariant mass lying in a $\pm50\mev$ window around the \Bpm-meson mass are removed. 
To reject background from \Dfourpi decays, the decay vertex of each long \KS candidate 
is required to be significantly displaced from the \D decay vertex along the beam direction.

The decay \BsDKst has a similar topology to \BdDKst, but exhibits much less 
\CP violation~\cite{LHCb-PAPER-2015-059}, since the decay
\decay{\Bs}{\Dz\Kstarzb} is doubly-Cabbibo suppressed compared to
\decay{\Bs}{\Dzb\Kstarzb}. These decays are used as a control channel in the invariant mass fit.
Background from partially reconstructed \BdsDstKst decays, where \Dstar stands for either \Dstarz or \Dstarzb,
are difficult to exclude since they have a topology very similar to the signal. 
The \DstDgam and \DstDpiz decays where the photon or the neutral pion is not reconstructed lead to
\BdsDKst candidates with a lower invariant mass than 
the \Bdors mass.

%% file: efficiency.tex
\section{Efficiency across the phase space}
\label{sec:efficiency}
The variation of the detection efficiency across the phase space is due to detector acceptance, 
trigger and selection criteria and PID effects. To evaluate this variation, a simulated sample generated uniformly
over the \DKSpipi phase space is used, after applying corrections for known differences between data and simulation 
that arise for the hardware trigger and PID requirements.

The trigger corrections are determined separately for two independent event categories. 
In the first category, events have at least one energy deposit in the hadronic calorimeter, associated with the 
signal decay, which passes the hardware trigger. In the second category, events are triggered only by 
particles present in the rest of the event, excluding the signal decay. The probability 
that a given energy deposit in the hadronic calorimeter passes the hardware trigger is evaluated with 
calibration samples, which are produced for kaons and pions separately, and give the trigger efficiency as a 
function of the dipole magnet polarity, the transverse 
energy and the hit position in the calorimeter. The efficiency functions obtained for the two categories
are combined according to their proportions in data.

The PID corrections are calculated with calibration samples of
\decay{\Dstarp}{\Dz\pip}, \decay{\Dz}{\Km\pip} decays. After background subtraction, the PID 
efficiencies for kaon and pion candidates are obtained as functions of momentum and pseudorapidity.
The product of the kaon and pion efficiencies, taking into account their correlation, gives the total
PID efficiency.

\begin{figure}[tb]
  \begin{center}
    \includegraphics[width=.49\textwidth]{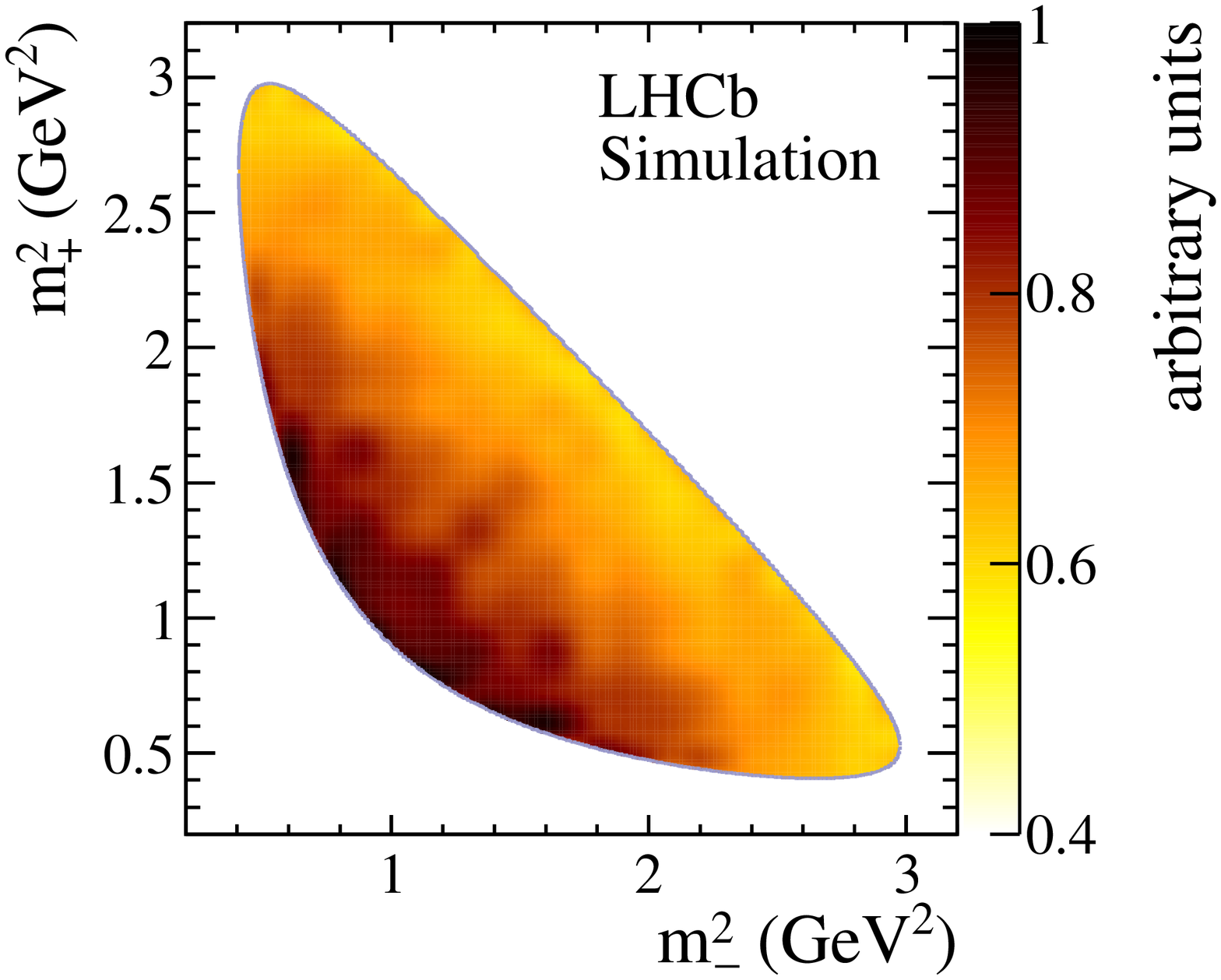}\includegraphics[width=.49\textwidth]{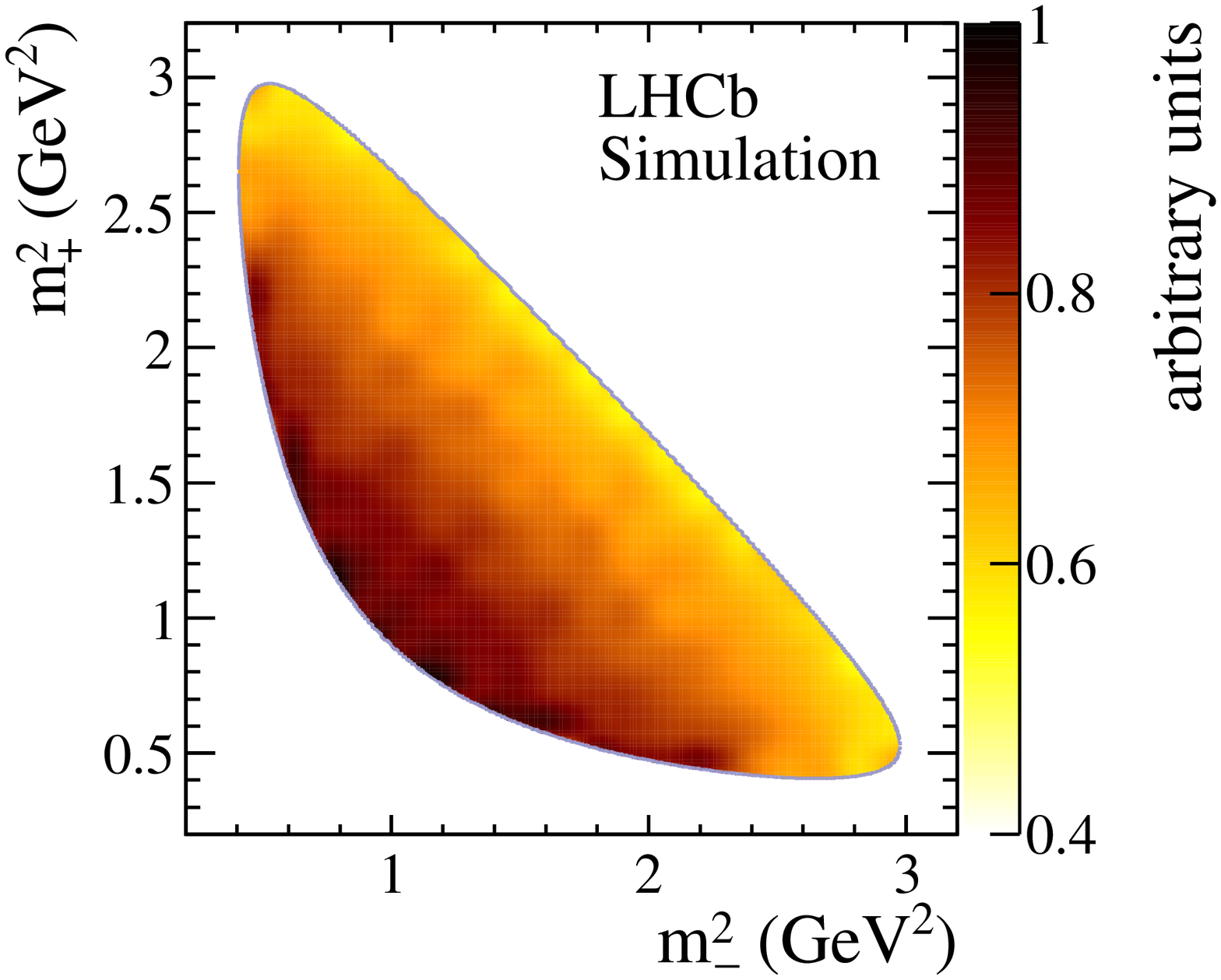}
  \end{center}
  \caption{Variation of signal efficiency across the phase space for (left) long and (right) downstream candidates.\label{fig:spline}}
\end{figure}

The various efficiency functions are combined to make two separate global efficiency functions, one for long candidates and one
for downstream candidates, which are used as inputs to the fit to obtain the Cartesian observables \zpm. 
To smooth out statistical fluctuations, an interpolation with a two-dimensional 
cubic spline function is performed to give a continuous description of
the efficiency $\eps(\mSqp,\mSqm)$, as shown in Fig.~\ref{fig:spline}.

%% file: fit.tex
\section{Analysis strategy and fit results}
\label{sec:fit}

To determine the \CP observables \zpm  defined in Eq.~\ref{eq:zpm}, an unbinned extended maximum likelihood
fit is performed in three variables: the \Bd candidate reconstructed invariant mass \mBd and the Dalitz variables
\mSqp and \mSqm. This fit is performed in two steps. First, the signal and background yields and some parameters of
the invariant mass PDFs are determined with a fit to the reconstructed \Bd invariant mass distribution, described in
Sec.~\ref{subsec:massfit}. An amplitude fit over the
phase space of the \D-meson decay is then performed to measure \zpm, using only
candidates lying in a $\pm25\mev$ window around the fitted \Bd mass, and taking the results of the invariant mass fit as 
inputs, as explained in Sec.~\ref{subsec:dalitzfit}. The \texttt{cfit} \cite{cfit}
library has been used to perform these fits.
Candidate events are divided into four subsamples, according to \KS\ type (long or downstream),
and whether the candidate is identified as a \Bd or \Bdb-meson decay.
In the \B-candidate invariant mass fit, the \Bd and \Bdb samples are combined, since identical distributions are 
expected for this variable, whilst in the \CP violation observables fit (\CP fit) they are kept separate.

\input{massfit}

\input{dalitzfit}

%% file: massfit.tex
\subsection{Invariant mass fit of \BdDKst candidates}
\label{subsec:massfit}
\begin{figure}[tb]
\begin{center}
\includegraphics[width=.8\textwidth]{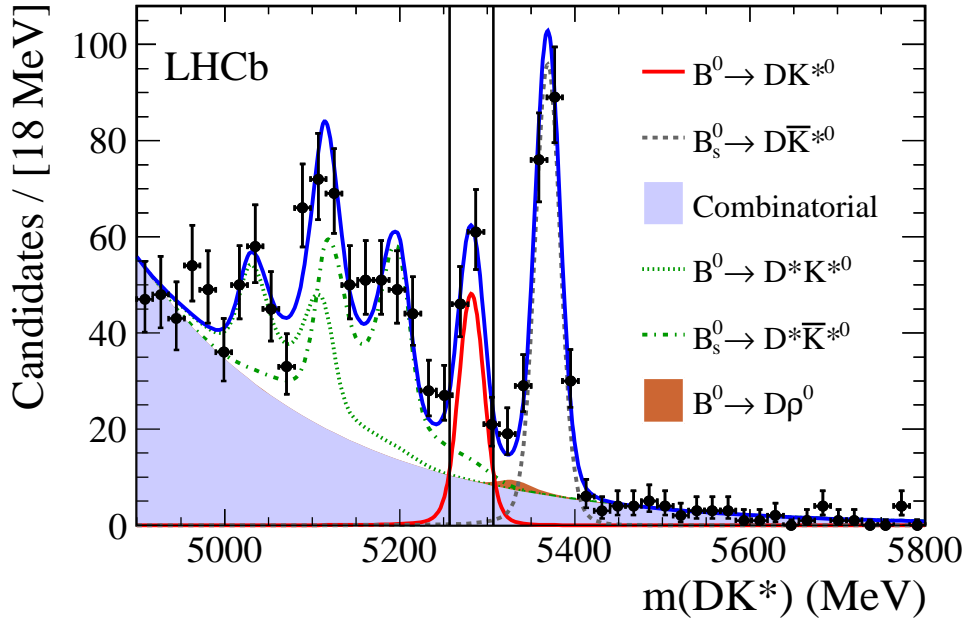}
\end{center}
\caption{Invariant mass distribution for \BdDKst long and downstream candidates. The fit result, 
including signal and background components, is superimposed (solid blue). The points are data, and 
the different fit components are given in the legend. The two vertical lines represent the signal
region in which the \CP fit is performed.\label{fig:massfit}}
\end{figure}

 An unbinned extended maximum likelihood fit to the reconstructed invariant mass distributions of the 
\Bd candidates in the range $[4900,5800]\mev$ determines the signal and background yields. 
The long and downstream subsamples are fitted simultaneously. The total PDF includes
several components: the \BdDKst signal PDF, background PDFs for \BsDKst decays, 
combinatorial background, partially reconstructed 
\BdsDstKst decays and misidentified \BdDrho decays, as illustrated in
Fig.~\ref{fig:massfit}.

The fit model is similar to that used in the analysis of \BdDKst decays
with \D-meson decays to two-body final states~\cite{LHCb-PAPER-2014-028}.
 The \BdDKst and \BsDKst components are each described as the sum of two 
Crystal Ball functions~\cite{Skwarnicki:1986xj} sharing the same central value, with the relative 
yields of the two functions and the tail parameters fixed from simulation. The separation between the central values of 
the \BdDKst and \BsDKst PDFs is fixed to the known \Bd-\Bs mass difference. 
The ratio of the \BdDKst and \BsDKst yields is constrained to be the same in both the long and downstream subsamples.
The combinatorial background is described with an exponential PDF. Partially 
reconstructed \BdsDstKst decays 
are described with non-parametric functions obtained by applying kernel density 
estimation~\cite{KDE} to distributions of simulated events. These distributions depend on the helicity
state of the \Dstarz meson.
Due to parity conservation in \DstDgam and \DstDpiz decays, two of the three helicity amplitudes
have the same invariant mass distribution. The \BsDstKst PDF is therefore a linear combination of two 
non-parametric functions, with the fraction of the longitudinal polarisation in the 
\BsDstKst decays unknown and accounted for with a free parameter in the fit.
 Each of the two functions describing the different helicity states is a weighted sum of 
non-parametric functions obtained from simulated \BsDstKstgam and \BsDstKstpiz decays, taking into account the 
known \DstDpiz and \DstDgam branching fractions~\cite{BES_DstarBR} and the appropriate efficiencies. The PDF for \BdDstKst decays 
is obtained from that for \BsDstKst decays, by applying a shift corresponding to the known \Bd-\Bs mass 
difference. In the nominal fit, the polarisation fraction is assumed to be the same for \BdDstKst and \BsDstKst decays.
The effect of this assumption is taken into account in the systematic uncertainties.
The \BdDrho component is also described with a non-parametric function obtained 
from the simulation, using a data-driven calibration to describe the pion-kaon misidentification efficiency. 
This component has a very low yield and, to improve the stability of the fit, a Gaussian constraint is applied, requiring the ratio of yields of \BdDrho
and \BsDKst to be consistent with its expected value.

The fitted distribution is shown in Fig~\ref{fig:massfit}. The resulting signal and background yields in a 
$\pm 25\mev$ range around the \Bd mass are given in Table~\ref{tab:yields}. This range corresponds to 
the signal region over which the \CP fit is performed.

\begin{table}[tb]
\caption{Signal and background yields in the signal region,
$\pm25\mev$ around the \Bd mass, obtained from the invariant mass fit.
Total yields, as well as separate yields for long and downstream candidates, are given.
\label{tab:yields}}
\begin{center}
\begin{tabular}{@{}lccc@{}}
Component          &   \multicolumn{3}{c}{Yield} \\
                   & Long            & Downstream      &    Total      \\
\midrule
\BdDKst            & $29\pm5\phantom{0}$        & $60\pm8\phantom{0}$        & $89\pm11$    \\[3mm]

\BsDKst            & $0.59\pm0.12$   & $1.21\pm0.23$   & $\phantom{0}1.8\pm0.3\phantom{0}$  \\
Combinatorial      & $\phantom{0}9.6\pm1.0\phantom{0}$     & $16.1\pm1.4\phantom{0}$    & $25.7\pm1.7\phantom{0}$ \\
\BdDstKst          & $0.06\pm0.02$   & $0.06\pm0.02$   & $0.12\pm0.03$\\
\BsDstKst          & $\phantom{0}4.1\pm0.8\phantom{0}$     & $\phantom{0}7.9\pm1.3\phantom{0}$     & $11.9\pm1.7\phantom{0}$ \\
\BdDrho            & $0.20\pm0.05$   & $0.37\pm0.09$   & $0.57\pm0.11$\\[3mm]

Total background   & $14.5\pm1.3\phantom{0}$    & $25.6\pm1.8\phantom{0}$    & $40.1\pm2.4\phantom{0}$ \\
\end{tabular}
\end{center}
\end{table}

%% file: dalitzfit.tex
\subsection{\CP fit}
\label{subsec:dalitzfit}
A simultaneous unbinned maximum likelihood fit to the four subsamples 
is performed
to determine the \CP violation observables \zpm.
The value of the coherence factor is fixed to the central value of $\kap = 0.958 ^{+0.005 +0.002}_{-0.010 -0.045}$, 
as measured in the recent \lhcb amplitude analysis of \BdDKPi decays~\cite{LHCb-PAPER-2015-059}.
The negative logarithm of the likelihood,
\begin{align}
\begin{aligned}
- \ln \mathcal{L} =
& - \sum_{\Bd  \mathrm{cand.}} \ln \left( \sum_{c} N_c f_c^{\mathrm{mass}}(m_B; \vec{q}_c^{\ \mathrm{mass}}) f_c^{\Bd\  \mathrm{model}}(\mSqp,\mSqm;\zpm,\kappa,\vec{q}_c^{\ \mathrm{model}}) \right)\\
& - \sum_{\Bdb \mathrm{cand.}} \ln \left( \sum_{c} N_c f_c^{\mathrm{mass}}(m_B; \vec{q}_c^{\ \mathrm{mass}}) f_c^{\Bdb\ \mathrm{model}}(\mSqp,\mSqm;\zpm,\kappa,\vec{q}_c^{\ \mathrm{model}}) \right)\\
& + \sum_c N_c,
\end{aligned}
\label{eq:likelihood}
\end{align}
is minimised,
where $c$ indexes the different signal and background components,
$N_c$ is the yield for each category, $f_c^\mathrm{mass}$ is the invariant mass PDF determined in the previous section, 
$\vec{q}_c^\mathrm{\ mass}$ are the mass PDF parameters, 
$f_c^{\B\ \mathrm{model}}$ is the amplitude PDF and 
$\vec{q}_c^\mathrm{\ model}$ are its parameters other than \zpm and \kap, which have been included explicitly. 

The non-uniformity of the selection efficiency over the \DKSpipi phase space is accounted 
for by including the function $\eps(\mSqp,\mSqm)$, introduced in Sec.~\ref{sec:efficiency},  within the  $f_c^{\B\ \mathrm{model}}$ PDF:
\begin{equation}
f_c^{\B\ \mathrm{model}}(\mSqp,\mSqm;z_\pm,\kappa,\vec{q}_c^{\ \mathrm{model}}) = \mathcal{F}_c(\mSqp,\mSqm;z_\pm,\kappa,\vec{q}_c^{\ \mathrm{model}})\ \eps(\mSqp,\mSqm),
\label{eq:Sig_DP_PDF}
\end{equation}
where $\mathcal{F}_c$ is the PDF of the amplitude model.

The model describing the amplitude of the \DKSpipi decay over the phase space, $A_f\left(m_+^2, m_-^2 \right)$, is identical to
that used previously by the \babar~\cite{delAmoSanchez:2010rq, delAmoSanchez:2010xz} 
and \lhcb~\cite{LHCb-PAPER-2014-017} collaborations.
An isobar model is used to describe 
$P$-wave (including $\rho(770)^0$, $\omega(782)$, Cabibbo-allowed and doubly Cabibbo-suppressed $\Kstar(892)^{\pm}$ and $\Kstar(1680)^-$)
and $D$-wave (including $f_2(1270)$ and $K^*_2(1430)^{\pm}$)
contributions.
The $\kaon \pi$ $S$-wave contribution ($K^*_0(1430)^{\pm}$) is described using a generalised LASS amplitude~\cite{LASS-Aston},
whilst the $\pi \pi$ $S$-wave contribution  is treated using a $P$-vector approach within the $K$-matrix formalism.
All parameters of the model are fixed in the fit to the values determined in 
Ref.~\cite{delAmoSanchez:2010xz}.\footnote{As previously noted in Ref.~\cite{LHCb-PAPER-2014-017}, 
  the model implemented by BaBar~\cite{delAmoSanchez:2010xz} differs from the formulation described
  therein. One of the two Blatt-Weisskopf coefficients was set to unity, and the imaginary part of the
  denominator of the Gounaris-Sakurai propagator used the mass of the resonant pair, instead of the mass
  associated with the resonance. The model used herein replicates these features without modification. It has
  been verified that changing the model to use an additional centrifugal barrier term and a modified
  Gounaris-Sakurai propagator has a negligible effect on the measurements.}

All components included in the fit of the \B-meson mass spectrum are included
in the fit for the \CP violation observables, with the exception of the \BdDstKst background, because its yield
within the signal region is negligible (Table~\ref{tab:yields}).
\CP violation is neglected for \BsDKst and \BsDstKst decays, 
since their Cabbibo-suppressed contributions are negligible.
The relevant PDFs are therefore
$\mathcal{F}_{\Bs \rightarrow D^{(*)} \Kstarzb}  = \mathcal{P}(\bar{A}_{f},0,0)$ and 
$\mathcal{F}_{\Bsb \rightarrow D^{(*)} \Kstarz} = \mathcal{P}(A_{f},0,0)$, where $\mathcal{P}$ is defined in Eq.~\ref{eq:functionalP}. 
For background arising from misidentified \BdDrho events, 
the \B flavour state cannot be determined, resulting in an incoherent sum 
of \Dz and \Dzb contributions:
$\mathcal{F}_{\Bd \rightarrow \D\rhoz}= ( |A_f|^2 + |\bar{A}_f|^2 ) / 2$.

\begin{figure}[tb]
\begin{center}
\includegraphics[width=.49\textwidth]{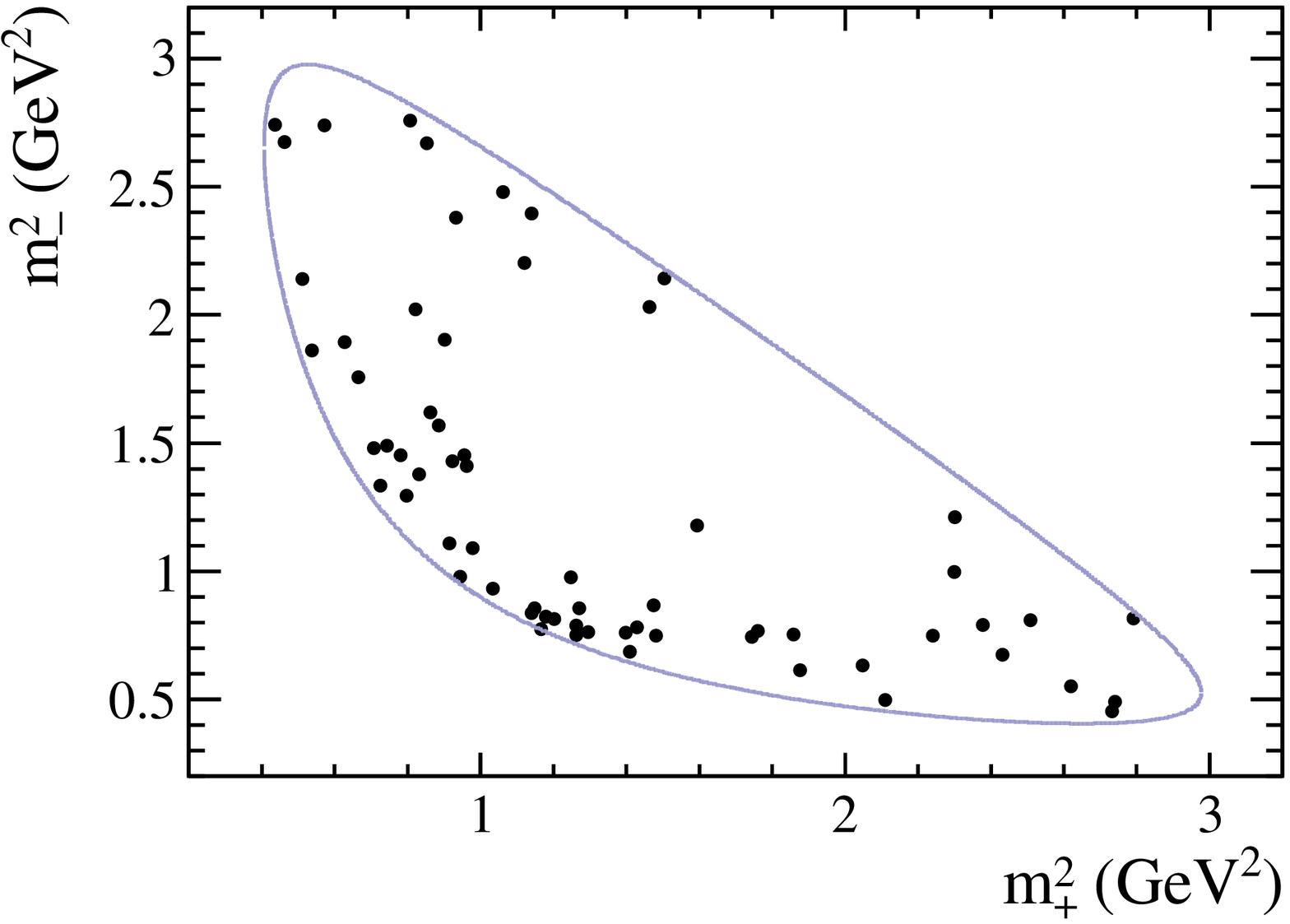}\put(-70,135){LHCb (a)}
\includegraphics[width=.49\textwidth]{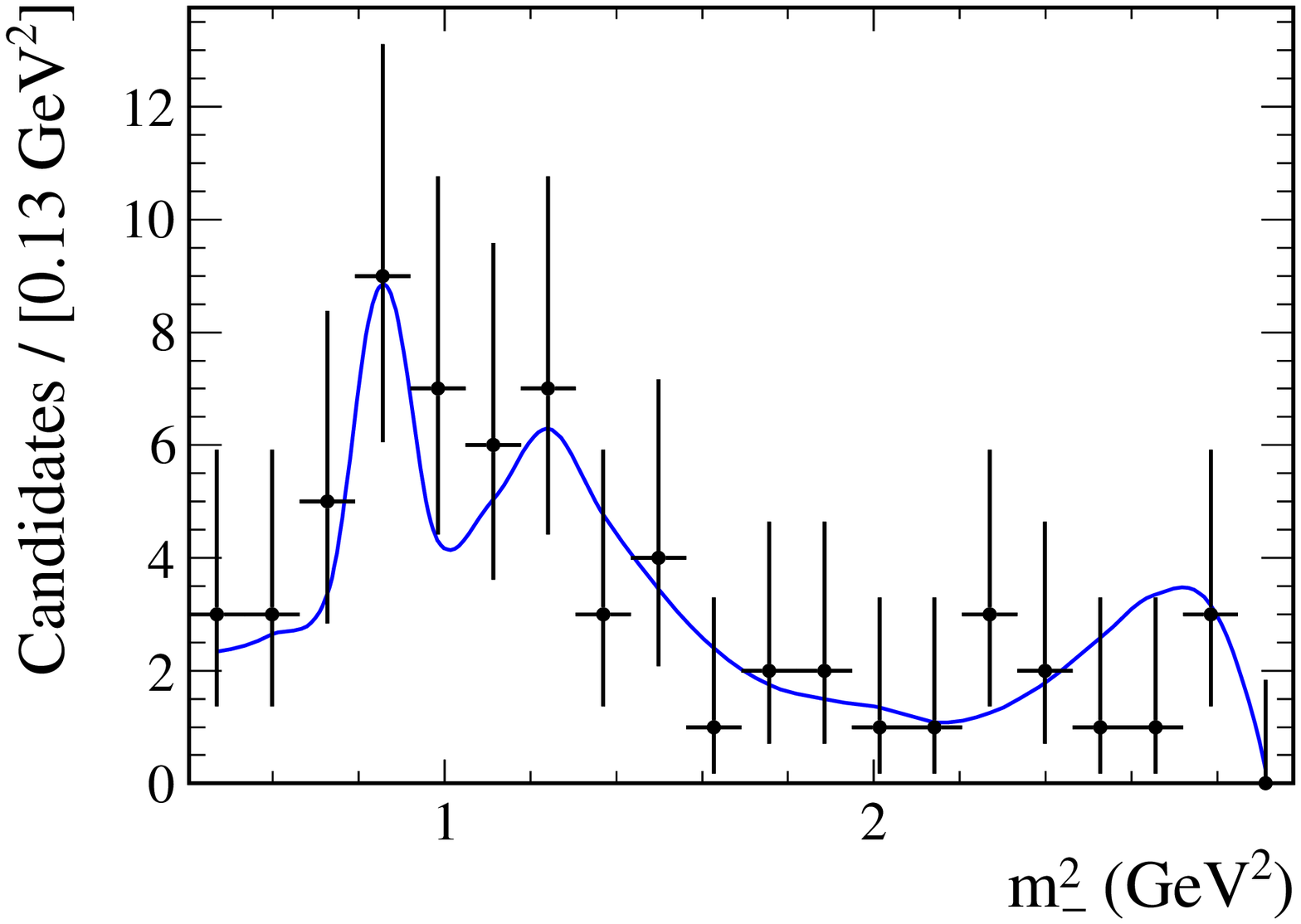}\put(-70,135){LHCb (b)}\\
\includegraphics[width=.49\textwidth]{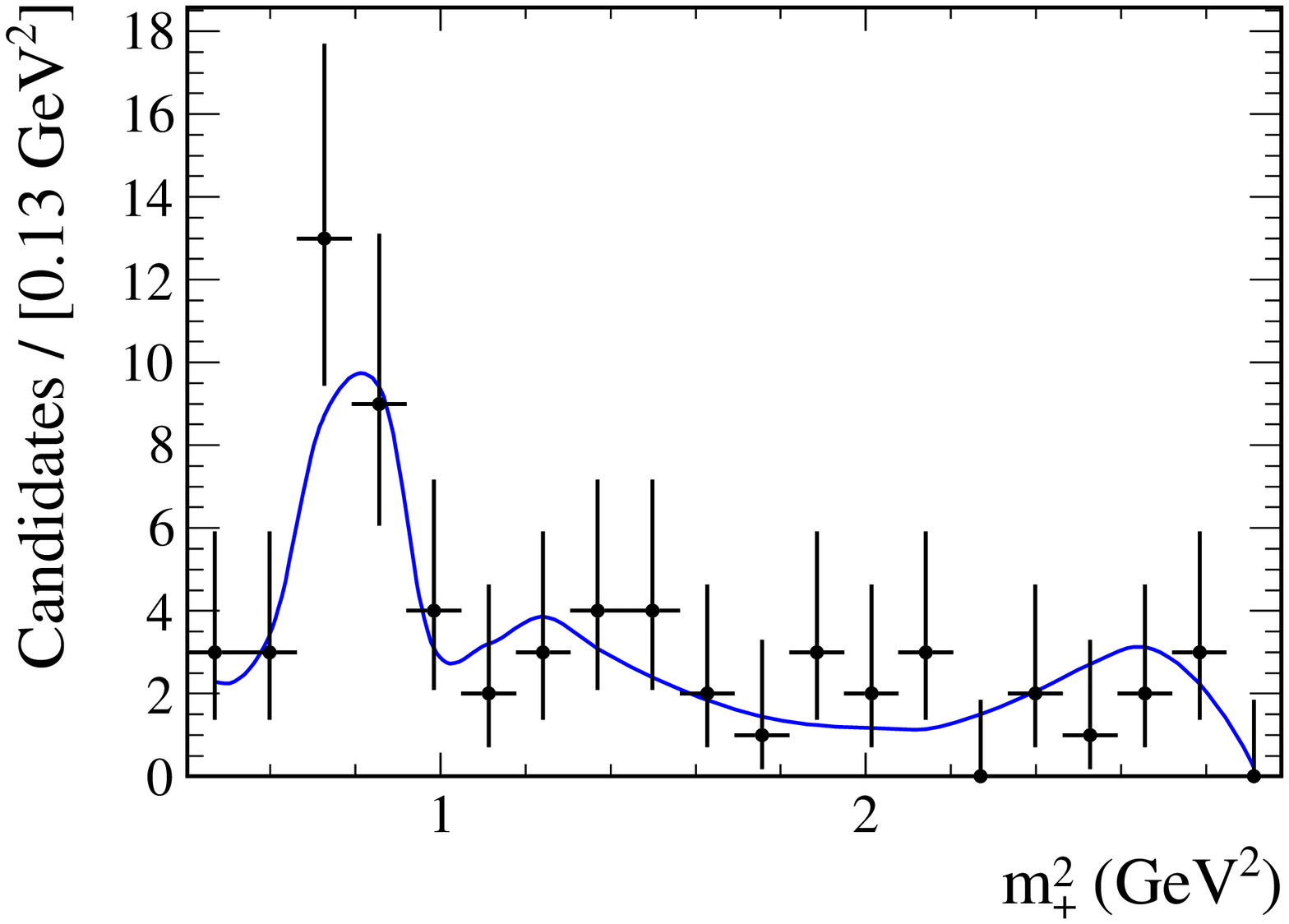}\put(-70,135){LHCb (c)}
\includegraphics[width=.49\textwidth]{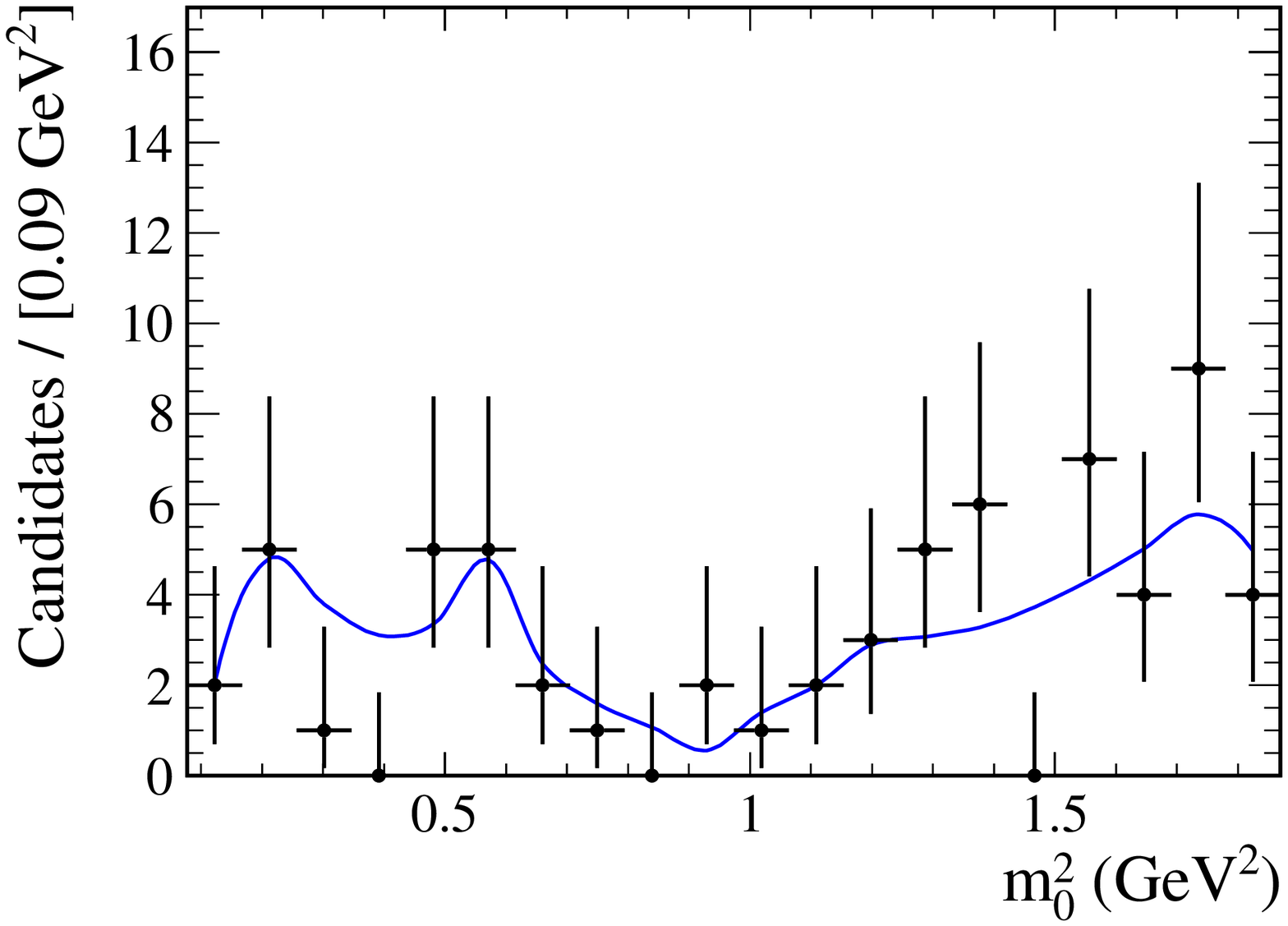}\put(-70,135){LHCb (d)}
\end{center}
\caption{Selected \BdDKst candidates, shown as (a) the Dalitz plot, and its projections on (b) \mSqm, (c) \mSqp and (d) \mSqz{}. The line superimposed on the projections corresponds to the fit result and the points are data.\label{fig:B0_data}}
\end{figure}

\begin{figure}[tb]
\begin{center}
\includegraphics[width=.49\textwidth]{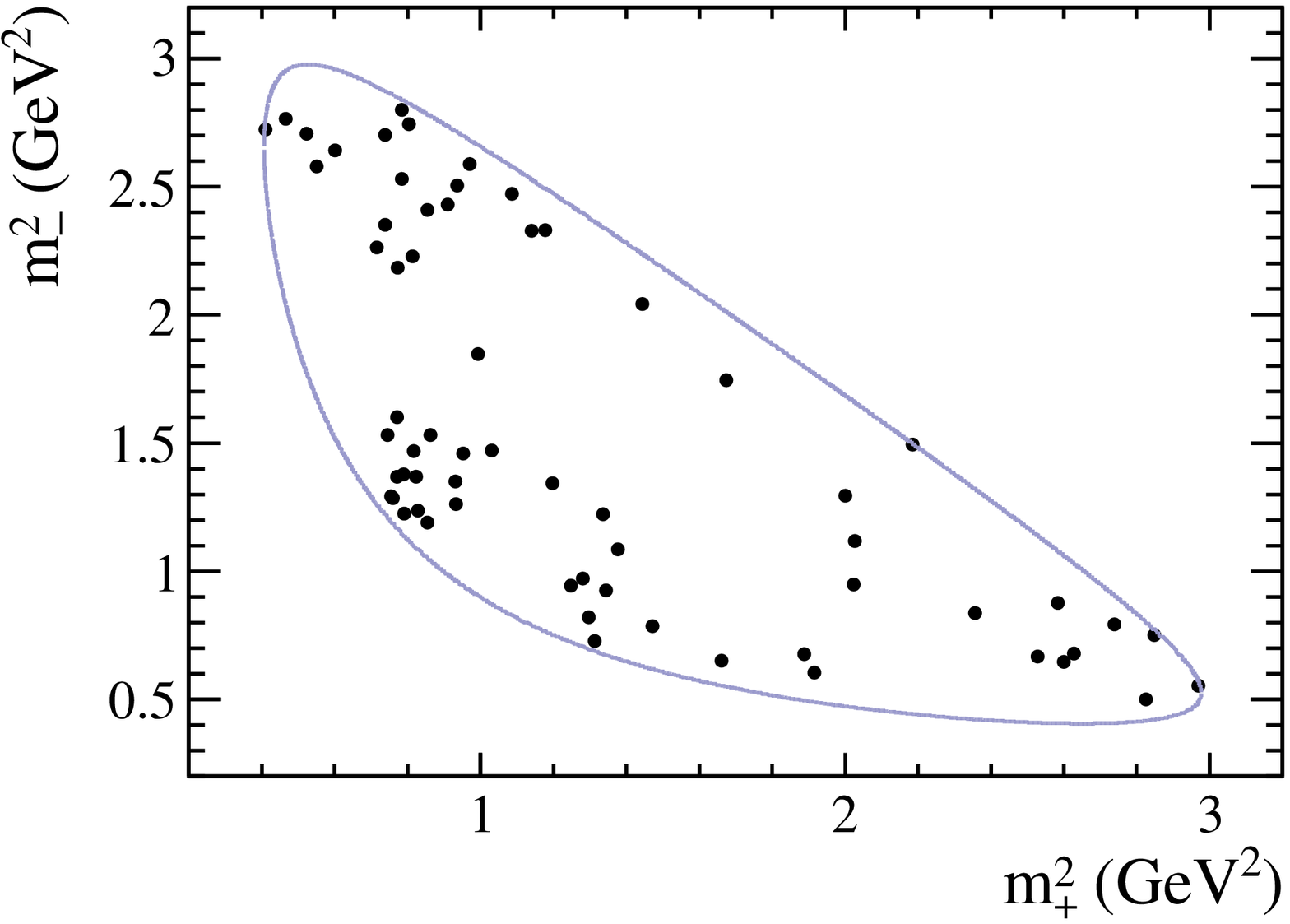}\put(-70,135){LHCb (a)}
\includegraphics[width=.49\textwidth]{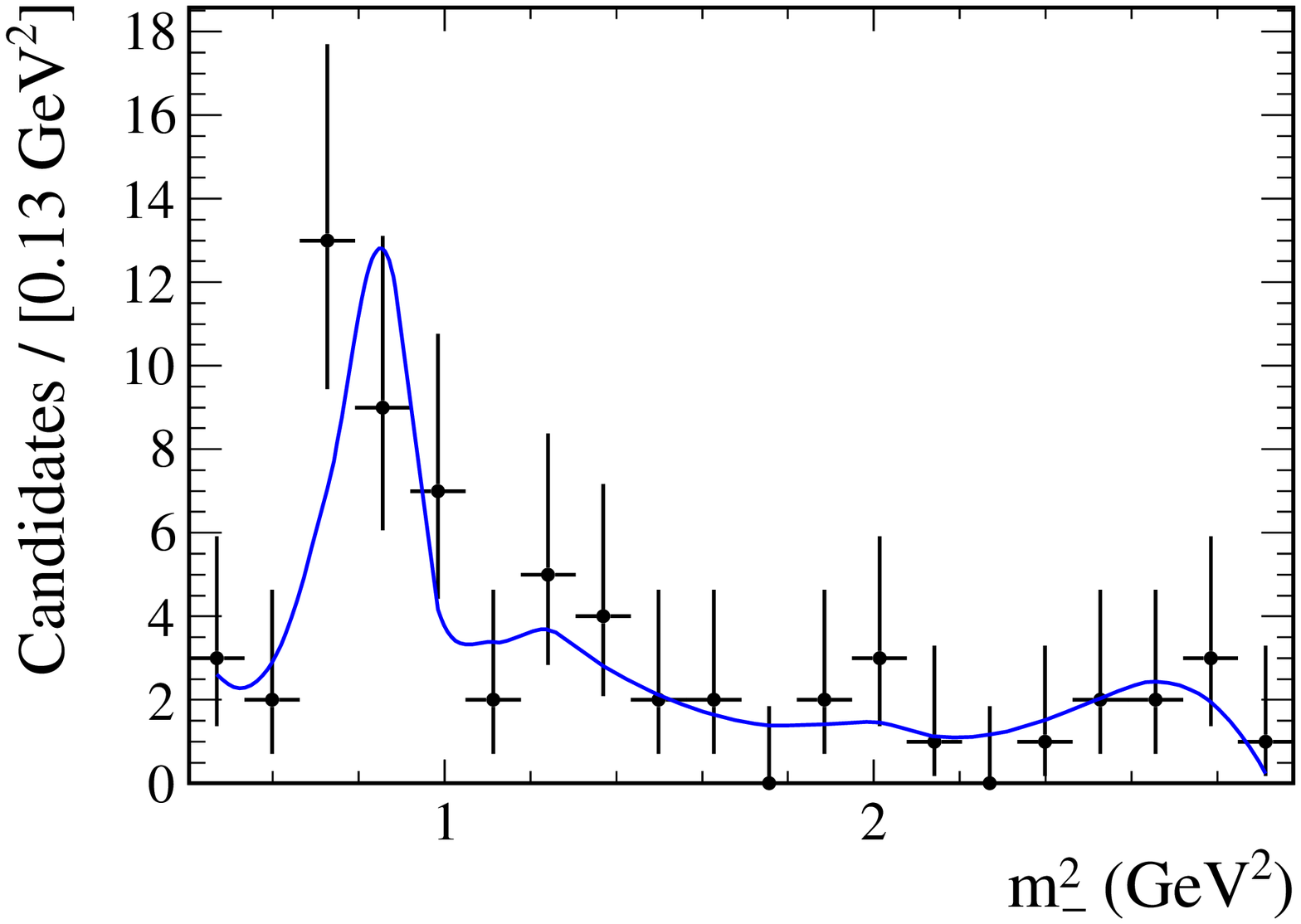}\put(-70,135){LHCb (b)}\\
\includegraphics[width=.49\textwidth]{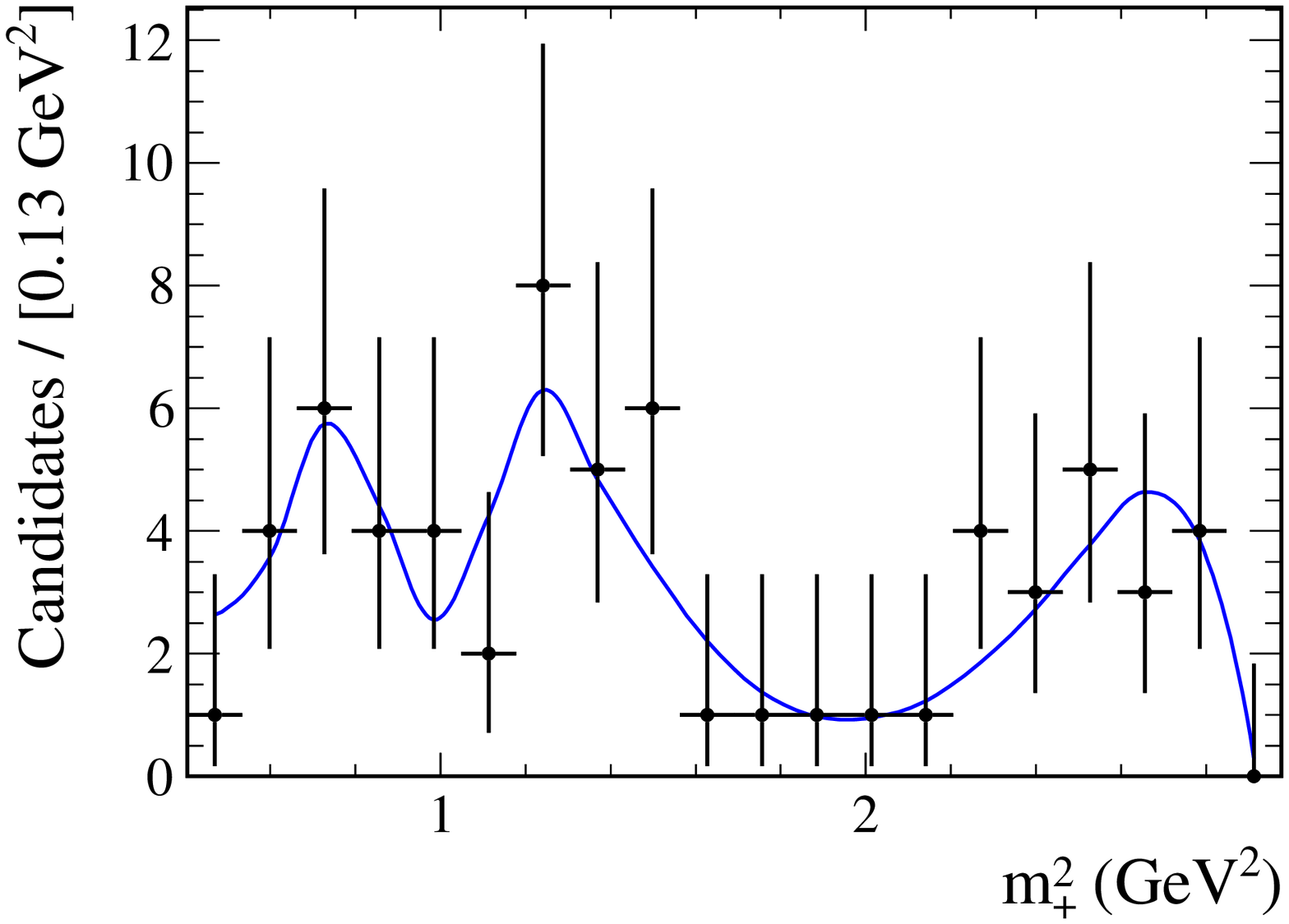}\put(-70,135){LHCb (c)}
\includegraphics[width=.49\textwidth]{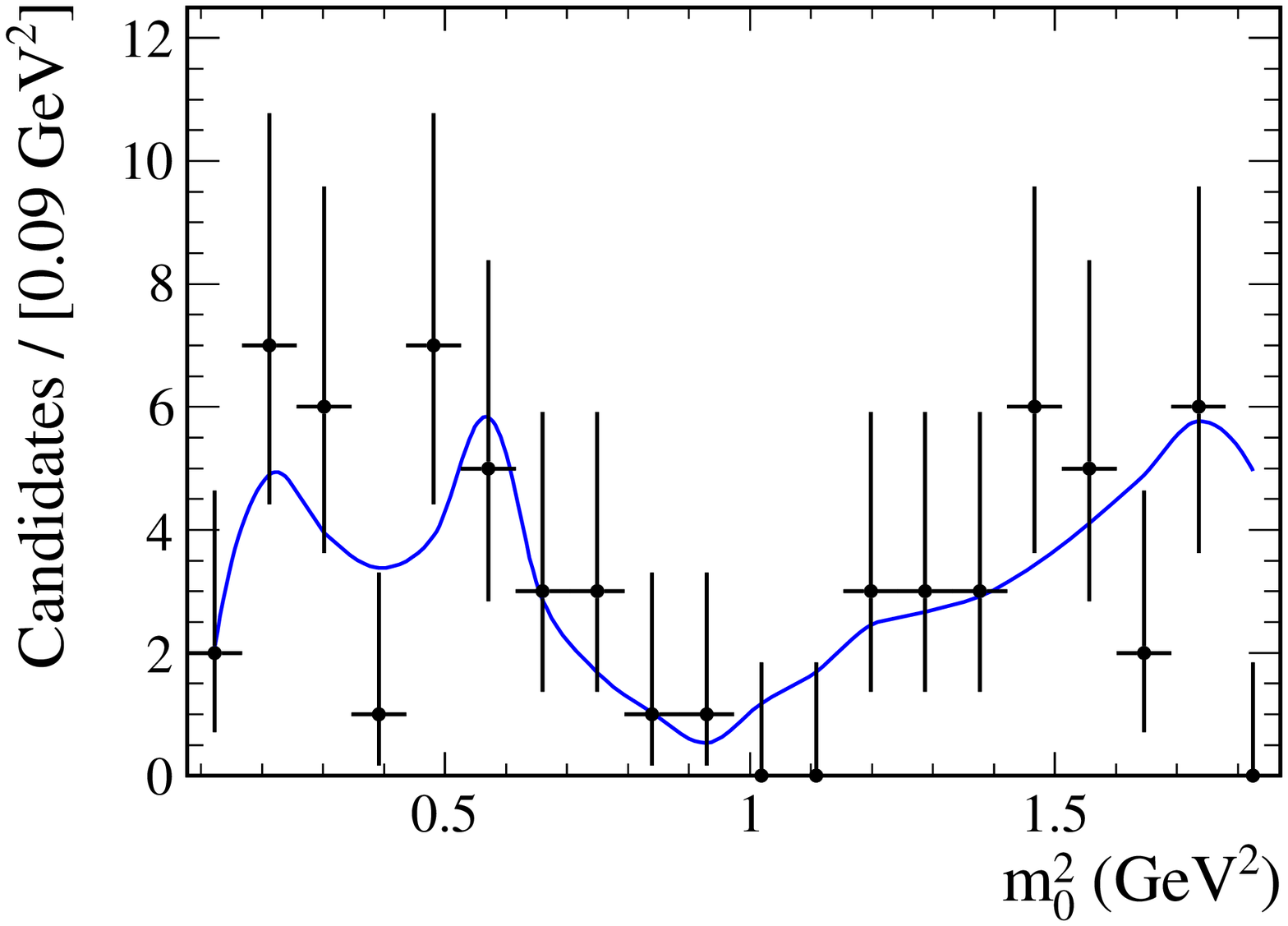}\put(-70,135){LHCb (d)}
\end{center}
\caption{Selected \BdbDKstb candidates, shown as (a) the Dalitz plot, and its projections on (b) \mSqm, (c) \mSqp and (d) \mSqz{}. The line superimposed on the projections corresponds to the fit result and the points are data.\label{fig:B0bar_data}}
\end{figure}

\begin{figure}[h]
  \begin{center}
    \includegraphics[width=0.50\textwidth]{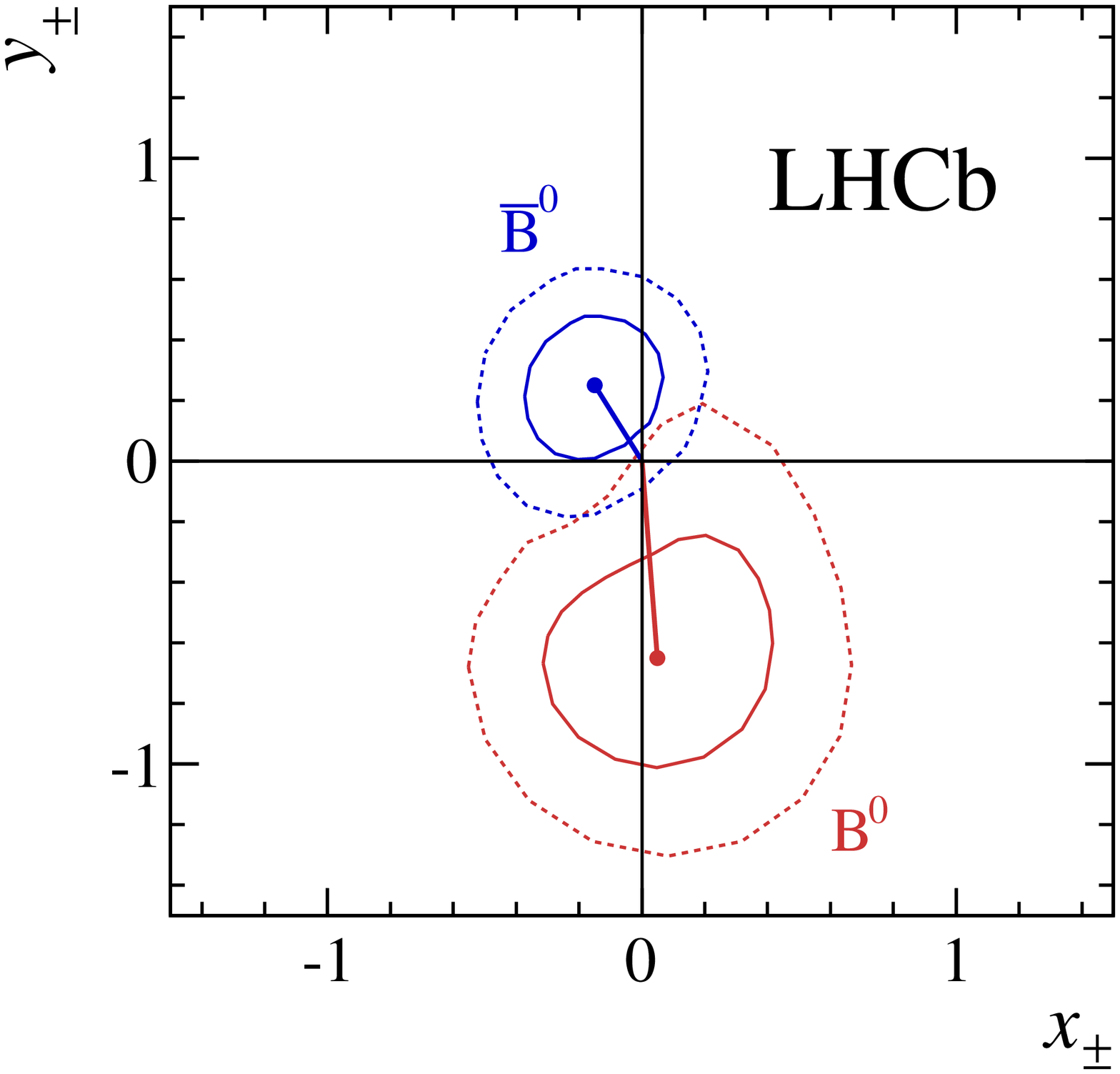}
    \caption{Likelihood contours at 68.3\% and 95.5\% confidence level for
      $(x_{+},y_{+})$\xspace (red) and $(x_{-},y_{-})$\xspace(blue), 
      obtained from the \CP fit.}
    \label{fig:cartesian_contours}
  \end{center}
\end{figure}

The combinatorial background is composed of two contributions:
one from non-$D$ candidates, and the other from real \D mesons combined with random tracks. 
Combinatorial \D candidates arise from random combinations of four charged tracks, incorrectly reconstructed as a \DKSpipi decay, 
and this contribution is assumed to be distributed uniformly over phase space, $\mathcal{F}_{\mathrm{Comb,\ non-}\D} = 1$, 
consistent with what is seen in the data.
Background from real \D candidates arises when the \Kst candidate is reconstructed from random tracks.
Consequently, the \B-meson flavour is unknown, resulting in an incoherent sum, 
$\mathcal{F}_{\mathrm{Comb,\ real}\ \D} = ( |A_f|^2 + |\bar{A}_f|^2 ) / 2$.
The relative proportions of non-\D and real \D meson backgrounds ($\order(30\%)$) are fixed using the results of a fit to the reconstructed invariant mass of the \D candidates in the signal \B mass region.
Figures~\ref{fig:B0_data} and \ref{fig:B0bar_data} show the Dalitz plot and its projections, 
with the fit result superimposed, for \Bd and \Bdb candidates, respectively.
A blinding procedure was used to obscure the values of the \CP parameters until all aspects of the analysis were finalised.
The measured values are
\begin{eqnarray*}
x_- &=& -0.15\, \pm0.14,\\
y_- &=& \phantom{-}0.25\, \pm0.15,\\
x_+ &=& \phantom{-}0.05\, \pm0.24,\\
y_+ &=& -0.65\, ^{~+0.24}_{~-0.23},
\end{eqnarray*}
where the uncertainty is statistical only.
The correlation matrix is
\begin{equation*}
\begin{matrix}
\begin{matrix} \;\; x_- \; & \;\; y_- \; & \;\; x_+ \; & \;\;  y_+ \end{matrix} \\ 
\begin{pmatrix} 
1      &  0.14   &  0     &  0    \\  
0.14   &  1      &  0     &  0    \\  
0      &  0      &  1     &  0.14 \\  
0      &  0      &  0.14  &  1    \\  
\end{pmatrix} \\
\end{matrix},
\label{eq:corrMatrix}
\end{equation*}
and the corresponding likelihood contours for \zpm
are shown in Fig.~\ref{fig:cartesian_contours}.

%% file: systematics.tex
\section{Systematic uncertainties}
\label{sec:systematics}

Several sources of systematic uncertainty on the evaluation of \zpm are considered, and are summarised in 
Table~\ref{tab:summary_syst}. 
Unless otherwise stated, for each source considered, the \CP fit is repeated and the 
differences in the \zpm values compared to the nominal results are taken as the systematic uncertainties.

\begin{table}[tb]
\caption{Summary of the systematic uncertainties on \zpm, in units of $(10^{-3})$. The total experimental 
and total model-related uncertainties are also given as percentages of the statistical uncertainties.\label{tab:summary_syst}}
\begin{center}

\begin{tabular}{@{}*5l@{}} 
Source of uncertainty & \dxm &  \dym   &  \dxp &  \dyp  \\
\hline

Efficiency & 5.4 & 1.1  & 11  & 1.8\\

Invariant mass fit              & 12 & 21   & 15  & 48\\

Migration over the phase space & 5.3 & 1.8 & 6.2 & 3.0\\
Misreconstructed signal        & 7.7 & 6.6 & 10 & 7.1\\

Background description &&&&\\
~~~~Non-\D background          &  20  & 15 & 28 & 47\\
~~~~Real \D background           &  0.1 & 0.4 & 0.2 & 1.0\\
~~~~\CP violation in \BsDstKst    &  1.5 & 0.8 &  4.0  & 1.6\\
~~~~\BuDpipipi contribution       &  0.6 & 1.4 &  0.8 & 2.3\\
~~~~\LbDppi contribution          &  0.1 & 0.7 &  0.5 & 1.6\\
\Kstar coherence factor (\kap)    &  4.8 & 2.4  &  8.5   & 2.6\\

\CP fit bias &  5 & 49 &  11   & 40  \\
\hline
Total experimental                & 26~(19\%)  & 56~(37\%) & 39~(16\%) & 78~(33\%) \\
\hline
Total model-related (see Table~\ref{tab:syst_model}) & \phantom{0}8~(5\%)    & \phantom{0}7~(5\%)      & 10~(4\%) & \phantom{0}5~(2\%)  \\
\end{tabular}
\end{center}
\end{table}

\input{syst_eff}

\input{syst_massfit}

The systematic uncertainty
due to the finite resolution in $\mSqpm$ is evaluated with a large number of pseudoexperiments. 
One nominal pseudodata sample is generated, with $z_\pm$ fixed to the values obtained from data. 
A large number of alternative samples are generated from the nominal one by smearing the $\mSqpm$ coordinates
of each event according to the resolution found in simulation and taking correlations into account.
For each \CP parameter, the width of the residual distribution is taken as the systematic uncertainty.

The misreconstruction of \BdDKst signal events is also studied.
This can occur \eg when the wrong final state pions of a real signal event are combined 
in the reconstruction of the $D$-meson candidate, leading to migration of this event within the $D$-decay phase space. 
The uncertainty corresponding to this effect is evaluated using pseudoexperiments. 
The effect of signal misreconstruction due to \Kstarz--\Kstarzb misidentification, corresponding to a $(\Kpm\pimp)\to(\pipm\Kmp)$ misidentification, 
is found to be negligible thanks to the PID requirements placed on the \Kstarz daughters.

\input{syst_bkg}

The \CP fit is verified using one thousand data-sized pseudoexperiments. In each experiment, the
signal and background yields, as well as the distributions used in the generation, are fixed to those found in data. The fitted values
of \zpm show biases smaller than the statistical uncertainties, and are included as systematic uncertainties. These biases are due to the current limited statistics and are found 
to reduce in pseudoexperiments generated with a larger sample size.

\input{syst_model}

The different systematic uncertainties are combined,
assuming that they are independent
 to obtain the total experimental uncertainties. Depending on the \xy parameters, the leading systematic 
uncertainties arise from the invariant mass fit, the description of the non-\D background and the fit biases. 
A larger data sample is expected to reduce all three of these uncertainties. Whilst not intrinsically statistical in nature, the  
systematic uncertainty due to the description of the non-\D background is presently evaluated using a conservative approach due to 
lack of statistics.
The total systematic uncertainties, including the model-related uncertainties, are significantly smaller
than the statistical uncertainties.

%% file: syst_eff.tex
The uncertainty on the description of the efficiency variation across the $D$-meson decay phase space arises
from several sources. 
Statistical uncertainties arise due to the limited sizes of the simulated samples used to determine the 
nominal efficiency function and of the calibration samples used to obtain the data-driven corrections to 
the PID and hardware trigger efficiencies.
Large numbers of alternative efficiency functions are created by smearing these quantities
according to their uncertainties.
For each fitted \CP parameter, the residual for a given alternative efficiency function is 
defined as the difference between its value obtained using this function, and that obtained 
in the nominal fit.  The width of the obtained distribution of residuals is taken as the 
corresponding systematic uncertainty.
Additionally, since the nominal fit is performed using an efficiency function obtained from the simulation 
applying only BDTA, the fit is repeated using an alternative efficiency function obtained using BDTB, 
and an uncertainty extracted.
The fit is also performed with alternative
efficiency functions obtained by varying the fraction of candidates triggered by at least one
product of the signal decay chain.
Finally, for a few variables used in the BDT, a small difference is observed between the 
simulation and the background-subtracted data sample. 
To account for this difference, the simulated events are reweighted to match the data, and 
the fit is repeated with the resulting efficiency function.

%% file: syst_massfit.tex
The $B$-meson invariant mass fit result is used to fix the fractions of signal and background and 
the parameters of the \Bd mass PDF shapes in the \CP fit. 
A large number of pseudoexperiments is generated, in which the 
free parameters of the invariant mass fit
are varied within their uncertainties, taking into account their correlations. 
The \CP fit is repeated for each variation.
For each \CP parameter, the width from a Gaussian fit to the resulting residual distribution
is taken as the associated systematic uncertainty.
This is the dominant contribution to the invariant mass fit systematic uncertainty quoted in Table~\ref{tab:summary_syst}.
Other uncertainties due to assumptions in the invariant mass fit  
are evaluated by allowing the \BdDKst/\BsDKst yield
ratio to be different for long and downstream categories, by varying the
\BdDrho/\BsDKst yield ratio, by varying the Crystal Ball PDF parameters
within their uncertainties and by testing alternatives to the Crystal Ball PDFs.
The proportions of \DstDgam and \DstDpiz in the \BdsDstKst background description are also varied, and
the effect of neglecting the \BdDstKst component in the \CP fit is evaluated.

%% file: syst_bkg.tex
The uncertainty arising from the background description 
is evaluated for several sources. 
The \CP fit is repeated with the fractions
of the two categories of combinatorial background (non-\D and real \D candidates)
varied within their 
uncertainties from the fit to the \D invariant mass distribution. 
Additionally, since in the nominal fit the non-\D candidates are assumed to be
uniformly distributed over the phase space of the \DKSpipi decay, 
the fit is repeated changing this contribution to the sum 
of a uniform distribution and a \Kstpm resonance. The
relative proportions of the two components are fixed 
based on the 
\mSqpm distributions found in data.
The fit is also repeated with the \D-meson decay model for the non-\D component set to 
the distribution of data in the \D mass sidebands. The uncertainty arising from the
poorly-known fraction of non-\D and real \D background
is the dominant systematic uncertainty for the \xpm parameters.

The description of the real \D combinatorial background assumes that the
probabilities of a \Dz or a \Dzb being present in an event are equal. The \CP violation observables
fit is repeated with the decay model for this background changed to include a 
\Dz--\Dzb production asymmetry, whose value is set to the measured \Dpm asymmetry
$(-1.0\pm0.3)\times10^{-2}$~\cite{LHCb-PAPER-2012-026}.

\CP violation is neglected in the \BsDstKst decay nominal description. The \CP fit is repeated with the inclusion
of a small component describing the suppressed decay amplitude of \decay{\Bs}{\Dstarz\Kstarzb}, 
with \CP violation parameters for this component fixed to $\gam=73.2\degrees$, 
$r_{\Bs}=0.02$ and $\delta_{\Bs}=\{0\degrees,45\degrees,90\degrees,135\degrees,180\degrees,225\degrees,270\degrees,315\degrees\}$. 
The model used to describe \BsDstKst decays consists of an incoherent sum of \DstDpiz and 
\DstDgam contributions. Between the \DstDpiz and \DstDgam decays, there is an effective strong 
phase shift of $\pi$ that is taken into account~\cite{Gamma_fromDst}.

The systematic uncertainties arising from the inclusion of background from misreconstructed \BuDpipipi and \LbDppi
decays are evaluated, by adding these components into the fit model. The \CP fit is also repeated 
with the \Kst coherence factor \kap varied within its uncertainty~\cite{LHCb-PAPER-2015-059}.

%% file: syst_model.tex
To evaluate the systematic uncertainty due to the choice of amplitude model for 
\DKSpipi,
one million \BdDKst\ and one million \BsDKst\ decays are simulated according to the nominal decay model, with
the Cartesian observables fixed to the nominal fit result.
These simulated decays are fitted with alternative models, each of which includes a single modification with respect
to the nominal model, as described in the next paragraph. 
Each of these alternative models is first used to fit the simulated \BsDKst decays to 
determine values for the resonance coefficients of the model. 
Those coefficients are then fixed in a second fit, to the simulated 
\BdDKst decays, to obtain \zpm.
The systematic uncertainties are taken to be the signed differences in the values of \zpm from the nominal
results.

\newcommand{\GeV}       {\ensuremath{\rm \, GeV}}

The following changes, labelled (a)-(u), are applied in the alternative models,
leading to the uncertainties shown in Table~\ref{tab:syst_model}: \vspace{-0.23cm}
\begin{itemize}\itemsep0pt \parskip0pt \parsep0pt \topsep0pt \partopsep0pt
  \item[$-$] $\pi\pi$ S-wave: The $F$-vector model is changed to use two other solutions of the $K$-matrix (from a total of three) determined from fits to
    scattering data~\cite{Anisovich:2002ij} (a), (b). The slowly varying part of the nonresonant term of the $P$-vector is removed (c).
  \item[$-$] $K  \pi$ S-wave: The generalised LASS parametrisation used to describe the
    $K^*_0(1430)^{\pm}$ resonance, is replaced by a relativistic Breit--Wigner propagator with parameters taken from Ref.~\cite{Aitala:2002kr} (d).
  \item[$-$] $\pi\pi$ P-wave: The Gounaris--Sakurai propagator is replaced by a relativistic Breit--Wigner propagator \cite{delAmoSanchez:2010xz,delAmoSanchez:2010rq} (e).
  \item[$-$] $K  \pi$ P-wave: The mass and width of the $\Kstar(1680)^-$ resonance are varied by their uncertainties from Ref.~\cite{LASS-Aston} (f)$-$(i).
  \item[$-$] $\pi\pi$ D-wave: The mass and width of the $f_2(1270)$ resonance are varied by their uncertainties from Ref.~\cite{PDG2014} (j)$-$(m).
  \item[$-$] $K  \pi$ D-wave: The mass and width of the $K^*_2(1430)^{\pm}$ resonance are varied by their uncertainties from Ref.~\cite{PDG2012} (n)$-$(q).
  \item[$-$] The radius of the Blatt--Weisskopf centrifugal barrier factors, $r_{{\rm BW}}$, is changed from $1.5\,\GeV^{-1}$ to $0.0\,\GeV^{-1}$ (r) and $3.0\,\GeV^{-1}$ (s).
  \item[$-$] Two further resonances, $\Kstar (1410)^{0}$ and $\rho(1450)$, parametrised with relativistic Breit--Wigner propagators,
    are included in the model \cite{delAmoSanchez:2010xz,delAmoSanchez:2010rq} (t).
  \item[$-$] The Zemach formalism used for the angular distribution of the decay products is replaced by the helicity formalism \cite{delAmoSanchez:2010xz,delAmoSanchez:2010rq} (u).
\end{itemize}
It results in total systematic uncertainties arising from the choice of amplitude model of
\begin{align*}
\delta x_- &= 8 \times10^{-3} , \\
\delta y_- &= 7 \times10^{-3} , \\
\delta x_+ &= 10 \times10^{-3} , \\
\delta y_+ &= 5 \times10^{-3} .
\end{align*}

\newcommand{\mrow}[2]   {\multirow{#1}{*}{#2}}
\newcommand{\fProd} {$f_{\pi\pi,j}^{\mathrm{pr}}$\xspace}
\newcommand{\kus}[1]{\mrow{4}{$\Kstar(#1)$\xspace}}
\newcommand{\kts}[1]{\mrow{4}{$K^*_2(#1)$\xspace}}
\newcommand{\ftw}[1]{\mrow{4}{$f_2(#1)$\xspace}}
\newcommand{\kstl}  {$\Kstar(1410)$\xspace}
\newcommand{\rhol}  {$\Prho(1450)$\xspace}
\newcommand{\BW}    {{\rm BW}}
\newcolumntype{R}{>{$}r<{$}}

\newcommand{\GS} {Gounaris-Sakurai}
\newcommand{\RBW}{relativistic Breit--Wigner}
\newcommand{\gLASS}{Generalised LASS}

\newcommand{\tr}[1]{\mrow{2}{#1}}
\newcommand{\TMT}       {(\times 10^{-3})}
\newcommand{\mcol}[2]   {\multicolumn{#1}{l}{#2}}
\newcommand{\mcolc}[2]  {\multicolumn{#1}{c}{#2}}
\newcommand{\mcolr}[2]  {\multicolumn{#1}{r}{#2}}

\begin{table}[tb]
  \begin{small}
  \caption{\small Model related systematic uncertainties for each alternative model, in units of $(10^{-3})$.  The relative signs
  indicate full correlation or anti-correlation.\label{tab:syst_model}}
  \begin{center}
  \begin{tabular}{@{}lll@{\hspace{35pt}}R@{\hspace{35pt}}R@{\hspace{35pt}}R@{\hspace{35pt}}R@{}}
             & \mcol{2}{Description}                 & \delta x_-      & \delta y_-      & \delta x_+      & \delta y_+      \\
    \midrule
    (a)      & \mcol{2}{$K$-matrix 1st solution}     &       -2        &        0.9      &        2        &        1        \\
    (b)      & \mcol{2}{$K$-matrix 2nd solution}     &        0.3      &        0.3      &        0.0      &       -0.5      \\
    \hline
    (c)      & \mcol{2}{Remove slowly varying}       &       -0.7      &        0.2      &        0.5      &        0.6      \\
             & \mcol{2}{part in $P$-vector}          &                 &                 &                 &                 \\
    \hline
    \tr{(d)} & \mcol{2}{\gLASS}                      &  \tr{$2$}       &  \tr{$3$}       &  \tr{$ -1$}     &  \tr{$ 3$}      \\
             & \mcol{2}{$\to$ \RBW}                  &                 &                 &                 &                 \\
    \tr{(e)} & \mcol{2}{\GS}                         &  \tr{$ 0.7$}    &  \tr{$0.0$}     &  \tr{$ -0.1$}   &  \tr{$ 0.8$}    \\
             & \mcol{2}{$\to$ \RBW}                  &                 &                 &                 &                 \\
    \hline
    (f)      & \kus{1680} & $m + \delta m$           &       -0.0      &        0.6      &        0.1      &        0.5      \\
    (g)      &            & $m - \delta m$           &       -0.2      &       -0.5      &        0.2      &       -0.9      \\
    (h)      &            & $\Gamma + \delta \Gamma$ &       -0.2      &        0.2      &        0.0      &       -0.2      \\
    (i)      &            & $\Gamma - \delta \Gamma$ &        0.2      &       -0.1      &        0.5      &       -0.2      \\
    \hline
    (j)      & \ftw{1270} & $m + \delta m$           &       -0.1      &        0.0      &        0.3      &       -0.2      \\
    (k)      &            & $m - \delta m$           &       -0.0      &        0.1      &        0.2      &       -0.2      \\
    (l)      &            & $\Gamma + \delta \Gamma$ &       -0.0      &        0.0      &        0.2      &       -0.2      \\
    (m)      &            & $\Gamma - \delta \Gamma$ &       -0.1      &        0.0      &        0.2      &       -0.2      \\
    \hline
    (n)      & \kts{1430} & $m + \delta m$           &        0.3      &        0.2      &        0.2      &       -0.2      \\
    (o)      &            & $m - \delta m$           &       -0.4      &       -0.2      &        0.3      &       -0.1      \\
    (p)      &            & $\Gamma + \delta \Gamma$ &       -0.2      &        0.2      &        0.1      &       -0.2      \\
    (q)      &            & $\Gamma - \delta \Gamma$ &        0.1      &       -0.1      &        0.3      &       -0.2      \\
    \hline
    (r)      & \mcol{2}{$r_{\BW} = 0.0 \GeV^{-1}$}   &       -2        &        0.7      &       -1        &       -0.3      \\
    (s)      & \mcol{2}{$r_{\BW} = 3.0 \GeV^{-1}$}   &        4        &       -2        &        4        &        2        \\
    (t)      & \mcol{2}{Add \kstl and \rhol}         &       -0.2      &       -0.2      &        0.3      &       -0.3      \\
    (u)      & \mcol{2}{Helicity formalism}          &       -6        &        6        &       -8        &        2        \\
    \hline
             & \mcol{2}{Total model related}    &  8 & 7 & 10 & 5 \\
  \end{tabular}
  \end{center}
  \end{small}
\end{table}

%% file: interpretation.tex
\section{Determination of the parameters \gam, \rbz and \deltabz}
\label{sec:interpretation}

\begin{figure}[htb]
\begin{center}
\includegraphics[width=.7\textwidth]{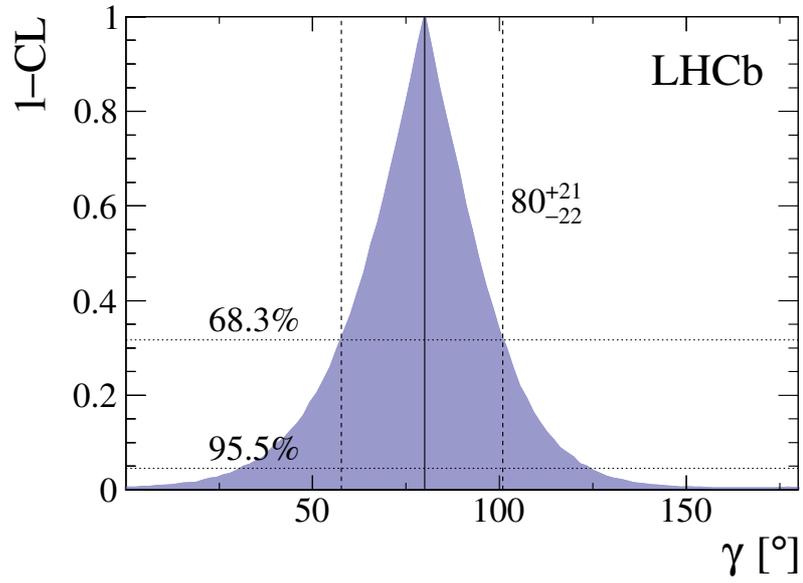}
\end{center}
\caption{Confidence level curve on \gam, obtained using the ``plugin'' method~\cite{Plugin}.\label{fig:CLcurve_gam}}
\end{figure}

\begin{figure}[htb]
\begin{center}
\includegraphics[width=.7\textwidth]{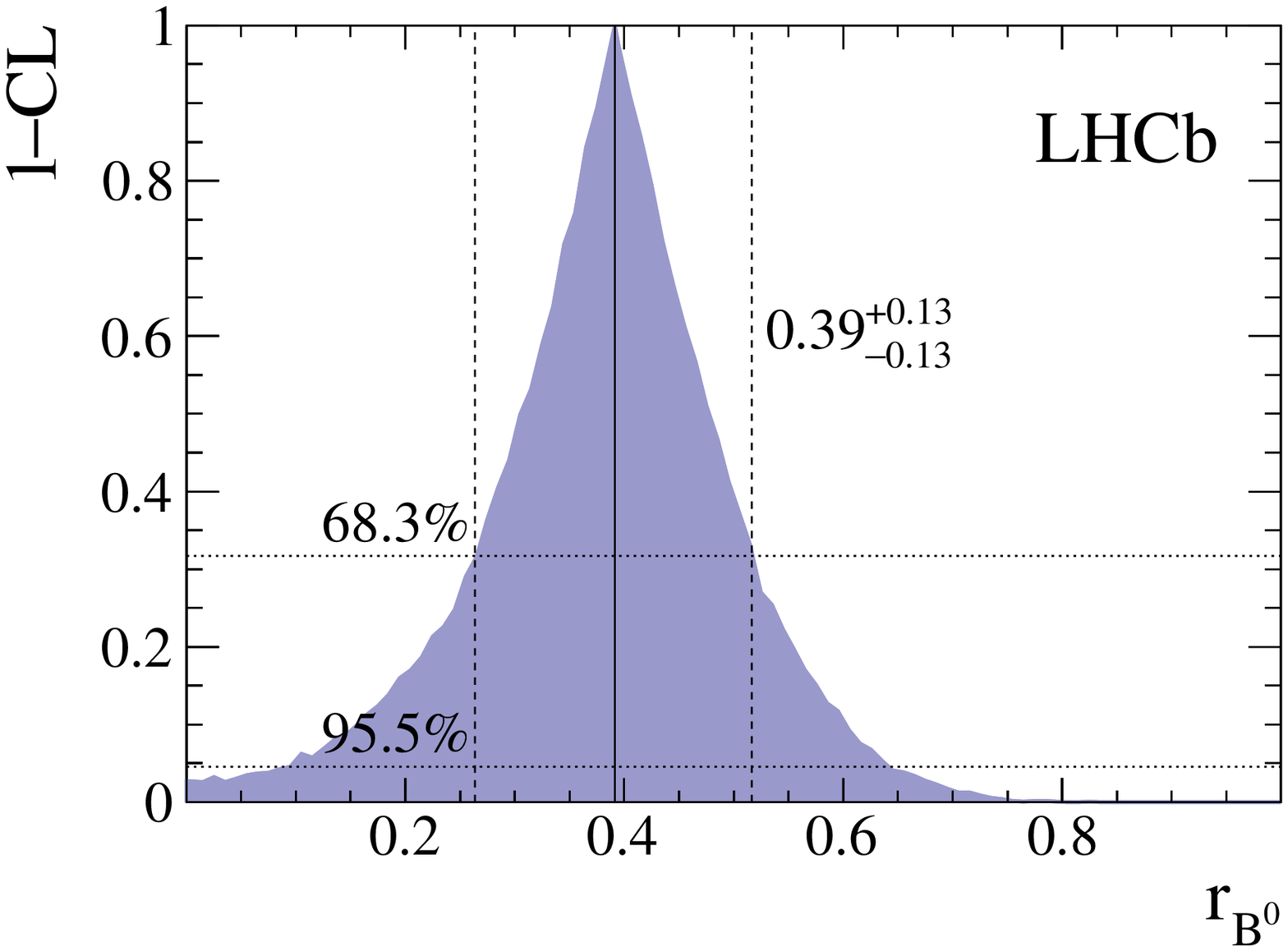}
\end{center}
\caption{Confidence level curve on \rbz, obtained using the ``plugin'' method~\cite{Plugin}.\label{fig:CLcurve_rbz}}
\end{figure}

\begin{figure}[htb]
\begin{center}
\includegraphics[width=.7\textwidth]{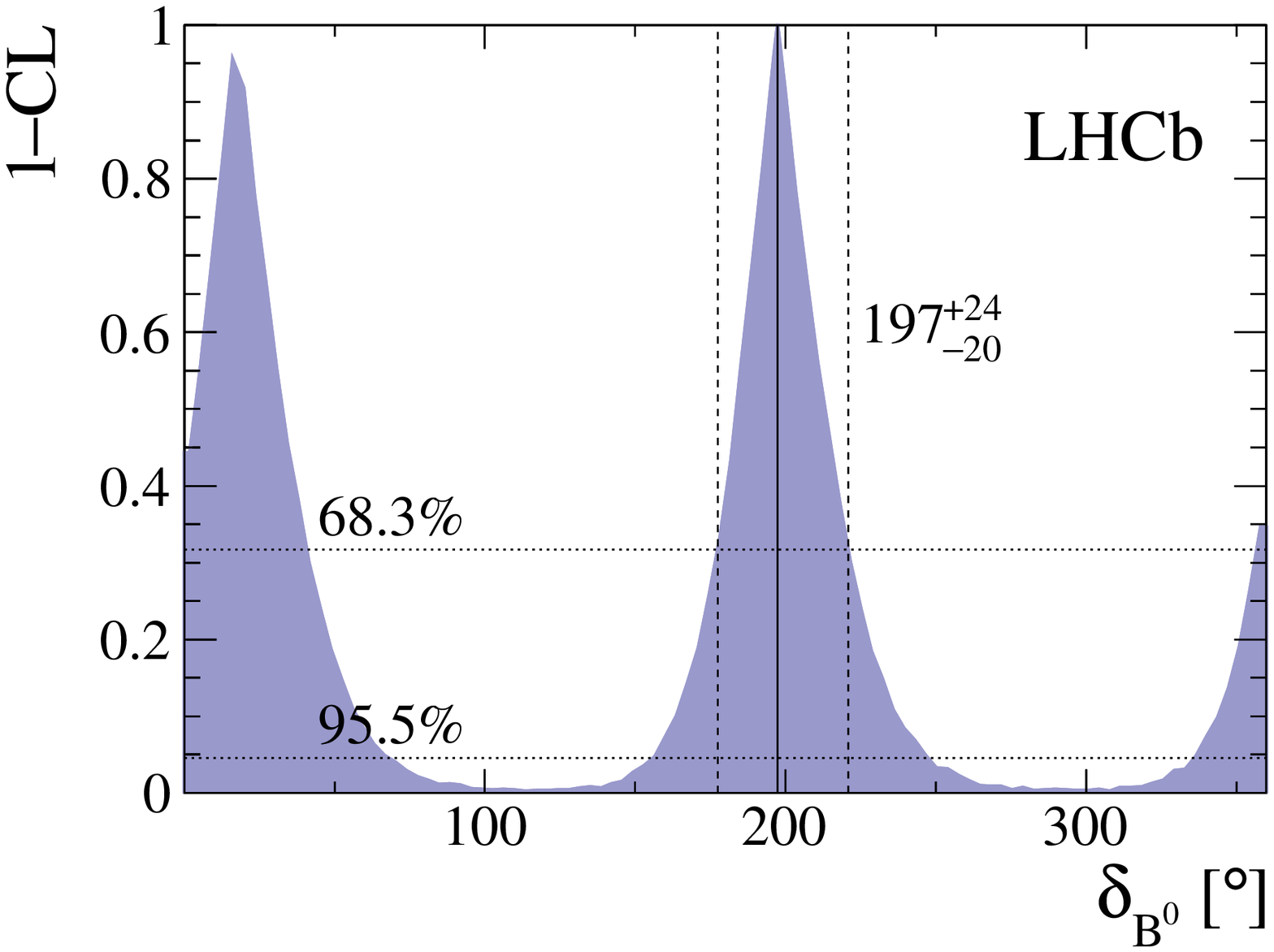}
\end{center}
\caption{Confidence level curve on \deltabz, obtained using the ``plugin'' method~\cite{Plugin}.
Only the \deltabz solution corresponding to $0<\gam<180\degrees$ is highlighted; the other maximum is due to the 
$(\deltabz,\gam)\to(\deltabz+\pi,\gam+\pi)$ ambiguity.\label{fig:CLcurve_deltabz}}
\end{figure}

\begin{figure}[htb]
\begin{center}
\includegraphics[width=.7\textwidth]{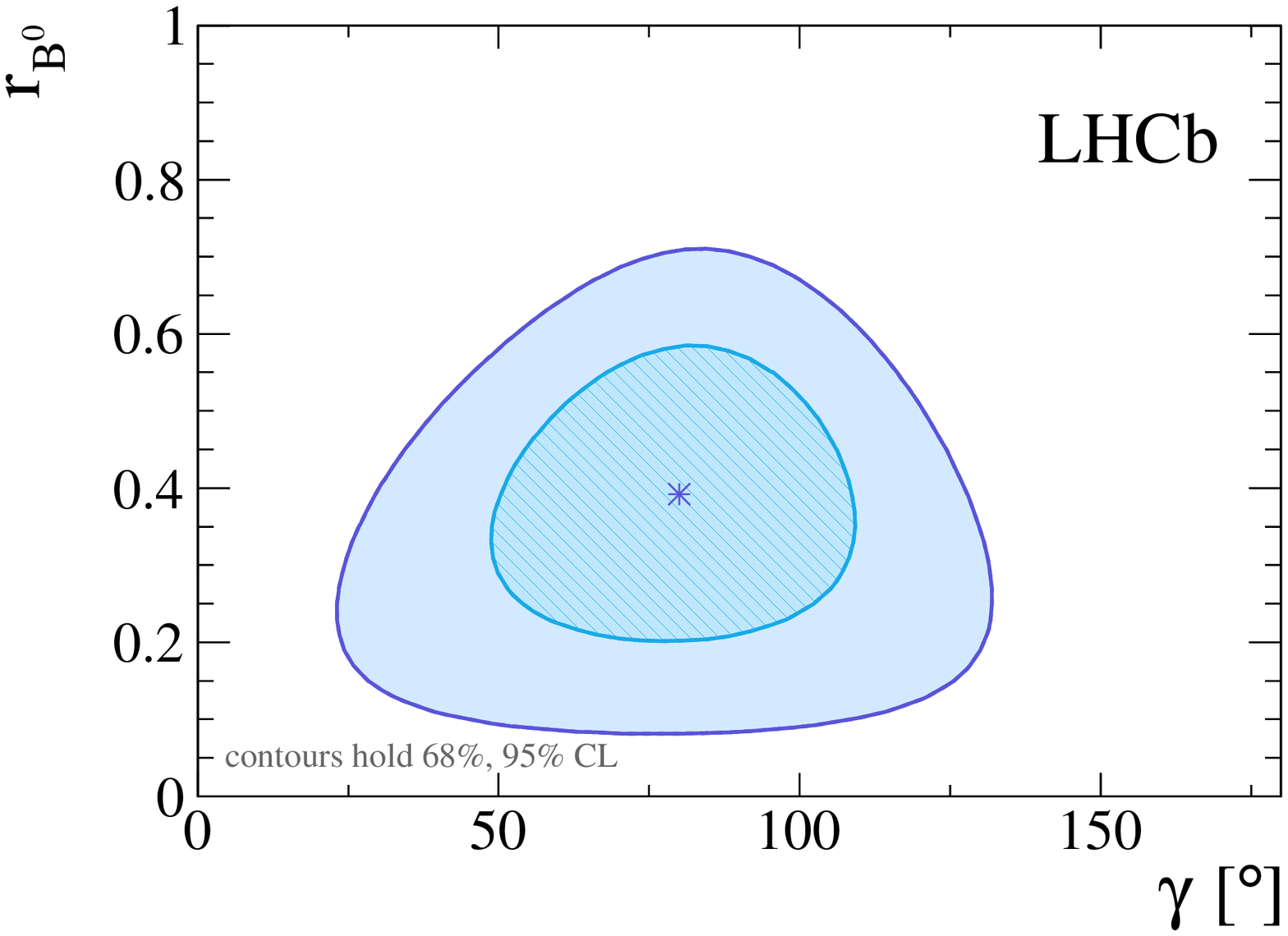}
\end{center}
\caption{Two-dimensional confidence level curves in the $(\gam,\rbz)$ plane, obtained using the profile-likelihood method. 
\label{fig:2dCLcurves_gr}}
\end{figure}

\begin{figure}[htb]
\begin{center}
\includegraphics[width=.7\textwidth]{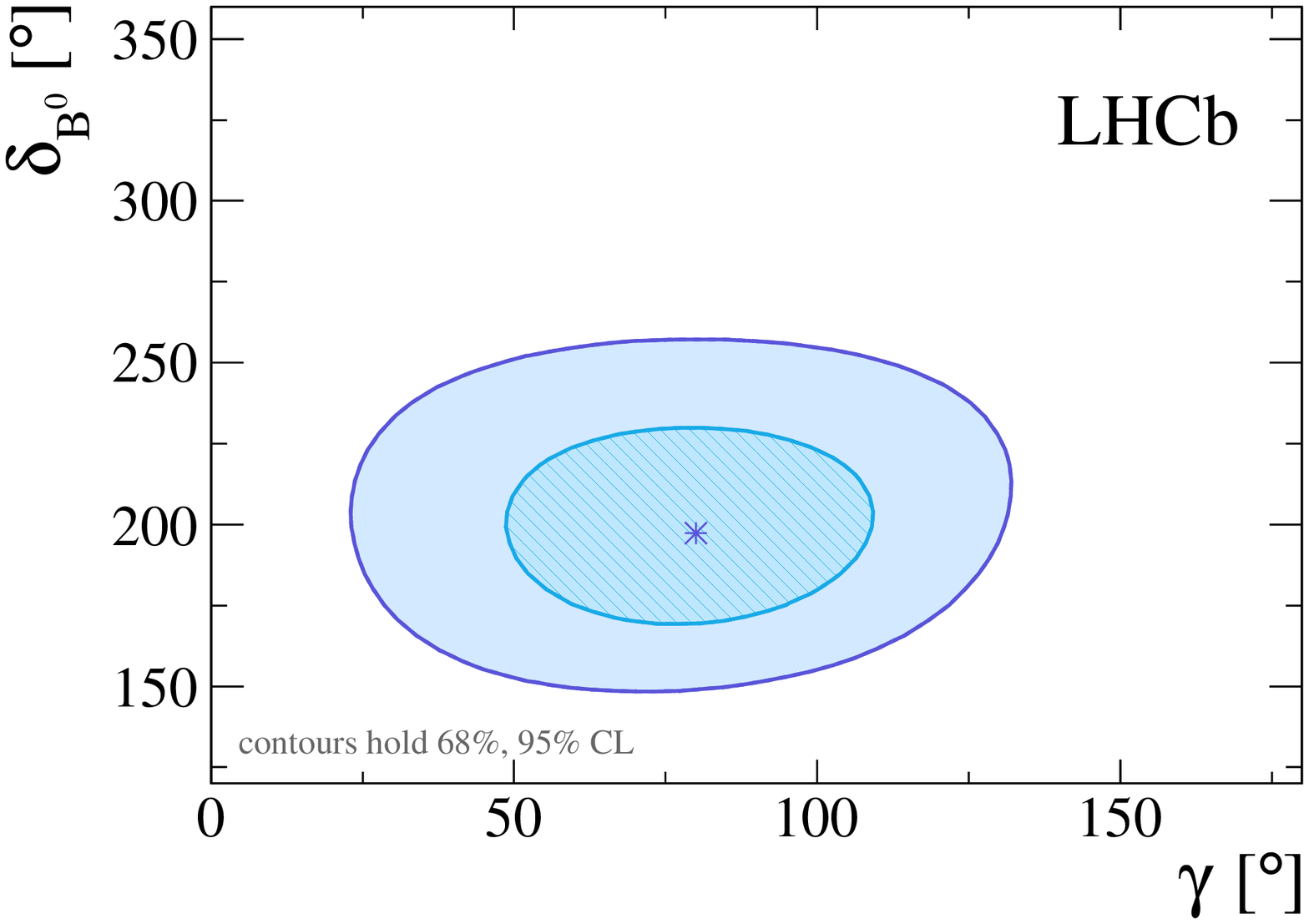}
\end{center}
\caption{Two-dimensional confidence level curves in the $(\gam,\deltabz)$ plane, obtained using the profile-likelihood method.\label{fig:2dCLcurves_gd}}
\end{figure}

To determine the physics parameters \rbz, \deltabz and \gam from the fitted Cartesian observables \zpm, the relations
\begin{align}                    
\begin{aligned}                    
\xpm &= \rbz \cos(\deltabz\pm\gam), \\
\ypm &= \rbz \sin(\deltabz\pm\gam),
\end{aligned}
\label{eq:xy_gam}
\end{align}
must be inverted. 
This is done using the \verb|GammaCombo| package, originally developed for the 
frequentist combination of \gam measurements by the \lhcb collaboration~\cite{LHCb-PAPER-2013-020,LHCb-CONF-2014-004}.
A global likelihood function is built, which gives the probability of observing a set of \zpm values given the true values \rbzdbzgam,
\begin{equation}
\mathcal{L}(\xm,\ym,\xp,\yp|\rbz,\deltabz,\gam). 
\end{equation}
All statistical and systematic uncertainties on \zpm are accounted for, as well as the statistical correlation between \zpm.
Since the precision of the
measurement is statistics dominated, correlations between the 
systematic uncertainties are ignored.
Central values for \rbzdbzgam are obtained by 
performing a scan of these parameters, to find the values that maximise
$\mathcal{L}(x_-^{\mathrm{obs}},y_-^{\mathrm{obs}},x_+^{\mathrm{obs}},y_+^{\mathrm{obs}}|\rbz,\deltabz,\gam)$,
where $z_\pm^{\mathrm{obs}}$ are the measured values of the Cartesian observables.
Associated confidence intervals may be obtained either from a simple profile-likelihood method,
or using the Feldman-Cousins approach~\cite{FeldmanCousins} combined with a ``plugin'' method~\cite{Plugin}.
Confidence level curves for \rbzdbzgam obtained using the latter method are shown in 
Figs.~\ref{fig:CLcurve_gam}, \ref{fig:CLcurve_rbz} and \ref{fig:CLcurve_deltabz}.
The measured values of \zpm are found to correspond to
\begin{align*}
\gamma &=\bigl(80^{+21}_{-22}\bigr)^{\circ}, \\
\rbz &=0.39\pm0.13, \\ 
\deltabz &= \bigl(197^{+24}_{-20} \bigr)^\circ.
\end{align*}
Intrinsic to the method used in this analysis~\cite{GGSZ}, there is a two-fold ambiguity in the solution; 
 the Standard Model solution $(0<\gam<180)\degrees$ is chosen. Two-dimensional confidence level curves obtained using the profile-likelihood method
are shown in Figs.~\ref{fig:2dCLcurves_gr} and \ref{fig:2dCLcurves_gd}.
 
%% file: conclusion.tex
\section{Conclusion}
\label{sec:conclu}

An amplitude analysis of \BdDKst decays, employing a model description of the \DKSpipi decay, 
has been performed using
data corresponding to an integrated luminosity of $3\invfb$,
recorded by \lhcb at a centre-of-mass energy of $7\tev$ in 2011 and $8\tev$ in 2012. The measured values of the \CP violation observables \mbox{$\xpm = \rbz \cos{(\deltabz \pm \gam)}$} and
\mbox{$\ypm = \rbz \sin{(\deltabz \pm \gam)}$} are
\begin{eqnarray*}
x_- &=& -0.15\, \pm 0.14 \pm 0.03 \pm 0.01,\\
y_- &=& \phantom{-}0.25\, \pm 0.15 \pm 0.06 \pm 0.01,\\
x_+ &=& \phantom{-}0.05\, \pm 0.24 \pm 0.04 \pm 0.01,\\
y_+ &=& -0.65~^{+0.24~~~}_{-0.23~~~} \pm 0.08 \pm 0.01,
\end{eqnarray*}
where the first uncertainties are statistical, the second are systematic and the third
are due to the choice of amplitude model used to describe the \DKSpipi decay.
These are the most precise measurements of these observables related to the neutral channel \BdDKst.
They place constraints on the magnitude of the ratio of the
interfering \B-meson decay amplitudes,
the strong phase difference between them
and the CKM angle \gam, giving
the values
\begin{eqnarray*}
 \gamma   &=& \bigl(80^{+21}_{-22}\bigr)^{\circ},\\
 \rbz     &=& 0.39\pm0.13,\\
 \deltabz &=& \bigl(197^{+24}_{-20}\bigr)^\circ.
\end{eqnarray*}
Here, \rbz and \deltabz are defined for a $K\pi$ mass region of $\pm50\mevcc$ around the \Kst mass and for an absolute value of
the cosine of the \Kstarz decay angle greater than $0.4$. These results are consistent with, and have lower total uncertainties 
than those reported in Ref.~\cite{LHCb-PAPER-2016-006}, where a model independent analysis method is used. The two results are 
based on the same data set and cannot be combined. The consistency shows that at the current level of statistical precision the 
assumptions used to obtain the present result are justified.

%% file: acknowledgements.tex
\section*{Acknowledgements}
\noindent We express our gratitude to our colleagues in the CERN
accelerator departments for the excellent performance of the LHC. We
thank the technical and administrative staff at the LHCb
institutes. We acknowledge support from CERN and from the national
agencies: CAPES, CNPq, FAPERJ and FINEP (Brazil); NSFC (China);
CNRS/IN2P3 (France); BMBF, DFG and MPG (Germany); INFN (Italy); 
FOM and NWO (The Netherlands); MNiSW and NCN (Poland); MEN/IFA (Romania); 
MinES and FANO (Russia); MinECo (Spain); SNSF and SER (Switzerland); 
NASU (Ukraine); STFC (United Kingdom); NSF (USA).
We acknowledge the computing resources that are provided by CERN, IN2P3 (France), KIT and DESY (Germany), INFN (Italy), SURF (The Netherlands), PIC (Spain), GridPP (United Kingdom), RRCKI and Yandex LLC (Russia), CSCS (Switzerland), IFIN-HH (Romania), CBPF (Brazil), PL-GRID (Poland) and OSC (USA). We are indebted to the communities behind the multiple open 
source software packages on which we depend.
Individual groups or members have received support from AvH Foundation (Germany),
EPLANET, Marie Sk\l{}odowska-Curie Actions and ERC (European Union), 
Conseil G\'{e}n\'{e}ral de Haute-Savoie, Labex ENIGMASS and OCEVU, 
R\'{e}gion Auvergne (France), RFBR and Yandex LLC (Russia), GVA, XuntaGal and GENCAT (Spain), Herchel Smith Fund, The Royal Society, Royal Commission for the Exhibition of 1851 and the Leverhulme Trust (United Kingdom).

%% file: LHCb_Authorship_flat_16-Feb-2016.tex
\centerline{\large\bf LHCb collaboration}
\begin{flushleft}
\small
R.~Aaij$^{39}$,
C.~Abell{\'a}n~Beteta$^{41}$,
B.~Adeva$^{38}$,
M.~Adinolfi$^{47}$,
Z.~Ajaltouni$^{5}$,
S.~Akar$^{6}$,
J.~Albrecht$^{10}$,
F.~Alessio$^{39}$,
M.~Alexander$^{52}$,
S.~Ali$^{42}$
% ...

{\footnotesize \it
% ...
$ ^{5}$Clermont Universit{\'e}, Universit{\'e} Blaise Pascal, CNRS/IN2P3, LPC, Clermont-Ferrand, France\\
$ ^{6}$CPPM, Aix-Marseille Universit{\'e}, CNRS/IN2P3, Marseille, France\\
% ...
$ ^{10}$Fakult{\"a}t Physik, Technische Universit{\"a}t Dortmund, Dortmund, Germany\\
% ...
$ ^{38}$Universidad de Santiago de Compostela, Santiago de Compostela, Spain\\
$ ^{39}$European Organization for Nuclear Research (CERN), Geneva, Switzerland\\
% ...
$ ^{41}$Physik-Institut, Universit{\"a}t Z{\"u}rich, Z{\"u}rich, Switzerland\\
$ ^{42}$Nikhef National Institute for Subatomic Physics, Amsterdam, The Netherlands\\
% ...
$ ^{47}$H.H. Wills Physics Laboratory, University of Bristol, Bristol, United Kingdom\\
% ...
$ ^{52}$School of Physics and Astronomy, University of Glasgow, Glasgow, United Kingdom\\
% ...
}
\end{flushleft}